\documentclass[%
reprint, 
superscriptaddress,
amsmath,amssymb,
aps, physrev,
pra,
]{revtex4-2}
\usepackage{float}
\usepackage{multirow}
\usepackage[utf8]{inputenc}
\usepackage{amssymb}   
\usepackage{braket}
\usepackage{graphicx}
\usepackage{dcolumn}
\usepackage{bm}
\usepackage[markup=underlined]{changes}

\begin{document}
	
	\title{Fast coherent control of a charge qubit on solid neon with a spin-qubit-compatible resonator}

	\author{Jun Wang}
	\email{jun.wang@riken.jp}
	\affiliation{RIKEN Center for Quantum Computing, 2-1 Hirosawa, Wako, Saitama, 351-0198, Japan}
	\author{Yiran Tian}
	\affiliation{RIKEN Center for Quantum Computing, 2-1 Hirosawa, Wako, Saitama, 351-0198, Japan}
	\author{Ivan Grytsenko}
	\affiliation{RIKEN Center for Quantum Computing, 2-1 Hirosawa, Wako, Saitama, 351-0198, Japan}
	\author{Asher Jennings}
	\affiliation{RIKEN Center for Quantum Computing, 2-1 Hirosawa, Wako, Saitama, 351-0198, Japan}
	\author{Beatriz P\'erez Gonz\'alez}
	\affiliation{Institute of Physics, University of Augsburg, Augsburg, 86159, Germany}
	\author{Xianjing Zhou}%
	\affiliation{Department of Mechanical Engineering, FAMU-FSU College of Engineering, Florida State University, Tallahassee, Florida 32310, USA}
	\author{Hirotaka Terai}%
	\affiliation{Advanced ICT Research Institute, National Institute of Information and Communications Technology (NICT), Kobe, 651-2492, Japan}
	\author{Dafei Jin}\affiliation{Department of Physics and Astronomy, University of Notre Dame, Notre Dame, Indiana 46556, USA}
	\author{Monica Benito}%
	\affiliation{Institute of Physics, University of Augsburg, Augsburg, 86159, Germany}
	\affiliation{Center for Advanced Analytics and Predictive Sciences, University of Augsburg, 86135 Augsburg, Germany}
	\author{Erika Kawakami}
	\email{e2006k@gmail.com}
	\affiliation{RIKEN Center for Quantum Computing, 2-1 Hirosawa, Wako, Saitama, 351-0198, Japan}
	\affiliation{RIKEN Pioneering Research Institute, 2-1 Hirosawa, Wako, Saitama, 351-0198, Japan.}

	\begin{abstract}
		Electrons floating in vacuum provide a clean platform for quantum information processing owing to their isolation from material defects. In particular, electrons on solid neon have emerged as a promising qubit platform because of their potentially long coherence times. Here, toward spin-qubit realization, we couple a single electron on solid neon to a magnetic-field-compatible superconducting NbTiN nanowire resonator. We realize a charge qubit and demonstrate microwave readout and coherent control, with Rabi frequencies up to 76 MHz, an order of magnitude larger than in previous studies. Under strong driving, we observe a qubit frequency shift from nonlinear interactions with the intense microwave field. Deterministic electron trapping at an intended position remains challenging due to solid neon surface roughness; we characterize the electron's position from its differential coupling to distinct electrodes. Although not trapped at an intended position, our estimates indicate that spin-qubit demonstrations remain feasible.
	\end{abstract}

	\maketitle
	
	\section{Introduction}
	Electrons floating above cryogenic substrates such as liquid helium and solid neon reside in vacuum and are therefore naturally isolated from solid-state defects, impurities, and other material-induced disorder. Electrons on helium have been the subject of extensive studies for several decades~\cite{Monarkha2004-un,Andrei1997Two-DimensionalSubstrates,Guo2025-zs}, leading to a well-established understanding of their physical properties and motivating numerous proposals that exploit their exceptionally pristine environment for qubit implementations~\cite{Platzman1999,Lea2000}, particularly spin qubits owing to their expected long coherence times~\cite{Lyon2006,Schuster2010,Kawakami2023-vf,Jennings2024-sb}. More recently, the strong coupling of a single trapped electron on helium to a superconducting resonator has been demonstrated, representing a major step toward qubit implementations based on electrons on helium, although an operational qubit has not yet been realized on liquid helium~\cite{Koolstra2019-mq,Koolstra2024-wc,Koolstra2025-qt}. In contrast, charge qubits based on electrons on solid neon have recently been realized, also using superconducting resonators, exhibiting coherence times reaching $T_2^*=$50~$\mu$s even for charge qubits, thereby attracting growing interest in this emerging platform~\cite{Zhou2022-nk,Zhou2023-iw,Li2025-em,Li2025-ci,Zou2026}. For spin qubits, even longer coherence times are theoretically expected~\cite{Chen2022-on}, with $T_2^*$ reaching $\sim 100~$s, and their experimental realization is highly anticipated.
	
	Realizing spin qubits using superconducting resonators requires resonators that remain operational in magnetic fields while maintaining a high quality factor. In this work, we pursue this goal using a superconducting NbTiN nanowire resonator. Nanowire resonators based on this material have been shown to remain relatively robust in magnetic fields~\cite{Samkharadze2016-xh,Samkharadze2018,Kroll2019-eo,Tian2025-fz}. Another advantage of NbTiN is its large kinetic inductance, which enables high characteristic impedance and enhanced zero-point voltage fluctuations, thereby increasing the coupling between the charge degree of freedom of the electron and microwave photons in the resonator. By coupling a single electron on solid neon to such a resonator, we realize a charge qubit in an architecture designed to be compatible with future spin-based implementations.
	
	A key challenge of the solid-neon platform is that the electron trapping landscape is strongly influenced by the microscopic morphology of the neon surface. Surface roughness can introduce uncontrolled electrostatic disorder, preventing deterministic positioning of electrons and making device behavior sensitive to the local trapping site~\cite{Kanai2024-bo,Zhou2022-nk}. To partially address this issue, we analyze the electron's coupling to multiple electrodes and thereby infer its likely position.
	
	Although the electron was not trapped at the intended position where charge–photon coupling is maximum, we nevertheless demonstrate microwave readout and coherent control of a single-electron charge qubit. We observe Rabi oscillations with frequencies up to $76~\mathrm{MHz}$, as well as Ramsey interference and Hahn-echo coherence. We measure $T_1$, $T_2^*$, and $T_2$. After annealing the neon film, we re-prepare the single-electron qubit and observe a longer relaxation time, which may be related to changes in the local environment of the electron, including possible variations in the neon film thickness. Under strong driving, we additionally observe a downward shift of the qubit transition frequency. We presumably attribute this nonlinear behavior to an AC Stark shift arising from the increased number of microwave photons in the resonator. 
	
	Furthermore, we theoretically consider the case where an electron is trapped at the deduced, albeit non-optimal, location and ferromagnets~\cite{Tokura2006,Pioro-Ladriere2008,Samkharadze2018,Tian2025-fz} are additionally integrated into the device. Using the charge-qubit parameters obtained in this work, we estimate that spin-qubit fidelities up to 99.995\% may be achievable in this platform, even without optimized electron positioning. These results establish NbTiN nanowire resonators on solid neon as a promising platform for realizing spin qubits.
	
	\section{Results}

	\begin{figure*}[t]
		\centering
		\includegraphics[width=\linewidth]{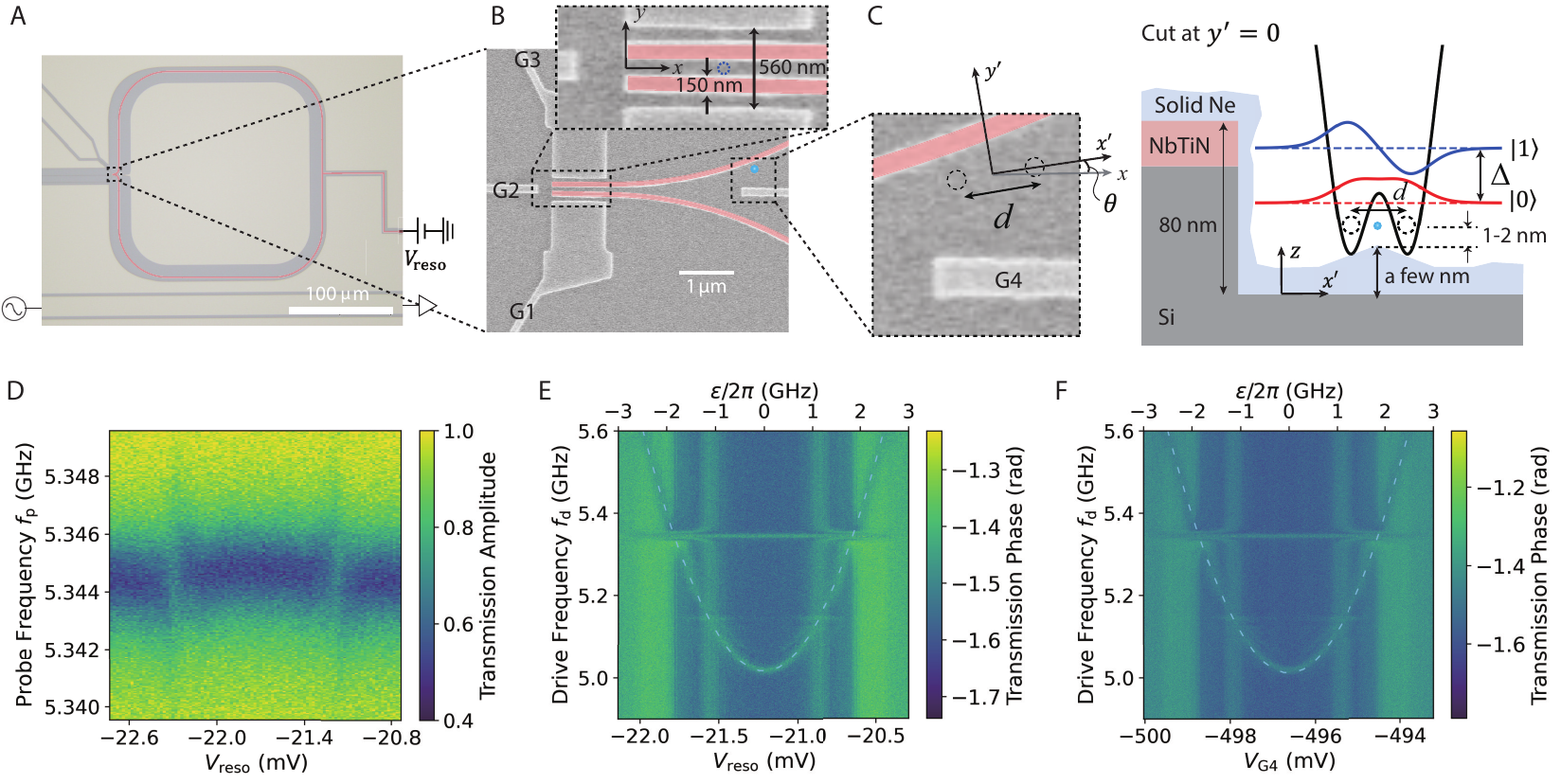}
		\caption{(A) Optical micrograph of the NbTiN nanowire resonator (false-color pink). The resonator is capacitively and inductively coupled to a feed line (bottom), through which microwave transmission is measured. The dotted square indicates the electron trapping region (enlarged in panel B). (B) Scanning electron micrograph of the electron trapping region, where the two ends of the nanowire resonator center conductor (highlighted in pink) approach each other with a 100 nm separation, with trapping electrodes G1--G4 located nearby. The light blue sphere denotes the electron position inferred from the experimental results (enlarged in panel C). Inset: Expanded view around the originally intended electron position, indicated by the blue dotted circle. 
			(C) Expanded view around the inferred electron position. The $x'$ axis connects the two potential minima, and $\theta$ denotes the angle between the $x$ and $x'$ axes. The right panel shows a cross-section of the device and potential along $y' = 0$.  After depositing solid neon (transparent blue) on the device, electrons are introduced and are possibly trapped by surface structures on the surface~\cite{Kanai2024-bo}. The double-well potential experienced by the electron at $\varepsilon = 0$ is illustrated by the black curve, while the red and blue curves represent the bonding and antibonding states, corresponding to the ground and first excited orbital states, i.e., the charge qubit states $|0\rangle$ and $|1\rangle$, respectively. Their energy splitting is denoted by $\Delta$. The positions of the two potential minima are indicated by the black dashed circles, and their separation is denoted by $d$.  (D) Transmission amplitude of the probe signal as a function of the probe frequency $f_p$ around $f_r$ and the resonator DC voltage $V_{\mathrm{reso}}$. Two avoided crossings appear at specific values of $V_{\mathrm{reso}}$, indicating coupling between the resonator and the electron orbital transition. (E,F) Two-tone spectroscopy. Transmission phase of the probe signal with $f_p = f_r$ while a drive signal is simultaneously applied and its frequency $f_d$ is swept, measured as a function of $V_{\mathrm{reso}}$ in (E) and the G4 DC voltage $V_{\mathrm{G4}}$ in (F). Both the drive and probe signals are applied to the feedline. Fits to Eqs.~\ref{eq:E}--\ref{eq:alpha_DC} are shown as dashed curves. In (E), $V_{\mathrm{G4}}=-500\,\mathrm{mV}$, while in (F), $V_{\mathrm{reso}}=-26.7\,\mathrm{mV}$. \label{fig1}}
	\end{figure*}

	\subsection{Device and single-electron trapping on solid neon}

	The device used in this work builds on the NbTiN superconducting nanowire resonator reported in Ref.~\onlinecite{Tian2025-fz}. We fabricate four DC electrodes (G1–G4) near the region where the two ends of the resonator approach each other (Fig.~\ref{fig1}A, B). The DC voltage applied to electrode $i$, denoted by $V_i$, where $i=\mathrm{G1},~\mathrm{G2},~\mathrm{G3},~\mathrm{G4}$ refers to the four DC electrodes and $i=\mathrm{reso}$ to the resonator electrode, tunes the electrostatic potential experienced by the trapped electron and thereby controls the energy levels of its orbital state. The resonator electrode can be DC-biased through a filter (see Supplementary Materials~\ref{sec:dc_filter_transmission}), while the resonator is AC-coupled to the feed line, through which both the microwave drive tone at frequency $f_d$ and the probe tone at frequency $f_p$ can be applied. The bare resonator exhibits a resonance at $f_r=5.34537$~GHz with internal and coupling quality factors
	$Q_i = 3583$ and $Q_c = 3785$, respectively, corresponding to a loaded quality factor
	$Q_l = 1841$ and a total decay rate $\kappa/2\pi = 2.9$~MHz. The relatively modest $Q$ value may be limited by losses associated with the DC bias circuitry and the capacitive coupling to the trapping electrodes.
	
	After deposition of neon and electrons~\cite{Tian2025-fz}, the electromagnetic environment of the resonator changes: the resonance shifts to $f_r = 5.34459$ GHz, with corresponding quality factors $Q_i = 2763$ and $Q_c = 3129$, yielding a loaded quality factor $Q_l = 1467$ and a total decay rate $\kappa/2\pi = 3.6$ MHz. To identify an electron exhibiting a clear resonant response, we sweep the DC voltages applied to electrodes G1--G4 and to the resonator while applying a microwave probe tone at $f_r$ to the feed line (Fig.~\ref{figS1}). When the orbital energy splitting of a trapped electron becomes resonant with the resonator, the transmission signal changes, allowing us to identify individual electrons. From these voltage scans, we select an electron that exhibits a clear and reproducible resonant response (see Supplementary Materials~\ref{sec:electron_selection_readout}).

	\subsection{Microwave spectroscopy}
	
	As discussed in Ref.~\onlinecite{Kanai2024-bo}, electrons on the neon surface can possibly be trapped by surface structures. The resulting electrostatic potential experienced by the electron can be described by a double-well potential (Fig.~\ref{fig1}C, Materials and Methods~\ref{sec:double_well_potential}). In the localized basis of the double-well potential, where the electron occupies either one or the other potential well, the qubit Hamiltonian reads~\cite{Hayashi2003}
	\begin{align}
		H=\frac{\hbar\varepsilon}{2}\sigma_z+\frac{\hbar\Delta}{2}\sigma_x .
	\end{align}
	where $\varepsilon$ is the energy detuning between the two localized states and $\Delta$ characterizes the coupling between the localized states in the two wells (Fig.~\ref{fig1}C). The orbital energy splitting is given by  
	\begin{align}
		E=\hbar\sqrt{\varepsilon^2+\Delta^2}=h f_q ,
		\label{eq:E}
	\end{align}
	where $f_q$ is the electron charge-qubit frequency.  The energy detuning $\varepsilon$ between the two wells is written as~\cite{VanderWiel2002}
	\begin{align}
		\hbar\varepsilon= \alpha_i^\mathrm{DC} e (V_i - V_i^{\mathrm{ss}}), \label{eq:alpha_DC}
	\end{align}
	where $V_i^{\mathrm{ss}}$ is the DC voltage applied to electrode $i$ at $\varepsilon=0$, and $\alpha_i^\mathrm{DC}$ characterizes the coupling strength between the electron and the DC voltage applied to electrode $i$.

	To characterize the qubit frequency as a function of the applied DC voltages, we perform microwave spectroscopy on the selected electron by sweeping the probe frequency $f_p$ around the resonator frequency $f_r$ while varying $V_{\mathrm{reso}}$. As predicted by Eqs.~\ref{eq:E} and \ref{eq:alpha_DC}, the orbital energy splitting matches the resonator frequency at two values of $V_{\mathrm{reso}}$, resulting in two resonances that are typically observed for a trapped electron (Fig.~\ref{fig1}D; Supplementary Materials \ref{sec:electron_selection_readout}). From Fig.~\ref{fig1}D, we extract the coupling strength between the electron orbital transition and the resonator to be \(g/2\pi = 2.1 \pm 0.2~\mathrm{MHz}\) (see Supplementary Materials~\ref{sec:extract_g} for the fitting procedure used to obtain \(g\)). The coupling strength can be expressed as~\cite{Ibberson2021-uq,Jennings2024-sb}
	\begin{equation}
		\hbar g = \alpha_{\mathrm{reso}}^{\mathrm{AC}} e V_{\mathrm{zpf}},
	\end{equation}
	where $\alpha_{\mathrm{reso}}^{\mathrm{AC}}$ characterizes the coupling between the electron and the AC voltage applied to the resonator electrode, and $V_{\mathrm{zpf}}$ is the zero-point voltage fluctuation of the resonator. With $L = 139~\mathrm{nH}$~\cite{Tian2025-fz}, the zero-point voltage fluctuation is $V_{\mathrm{zpf}} = \sqrt{\frac{\hbar (2\pi f_r)^3 L}{2}} = 16.7~\mu\mathrm{V}$. This yields $\alpha_{\mathrm{reso}}^{\mathrm{AC}} = (5.2 \pm 0.5) \times 10^{-4}$.

	To map the qubit frequency over a broader range around the resonator frequency, we perform two-tone spectroscopy. In this measurement, the probe tone is fixed at the resonator frequency ($f_p=f_r$) to monitor the resonator response, while a drive tone $f_d$ is swept and the DC voltage applied to electrode $i$ is varied. Figures~\ref{fig1}E and F show two-tone spectroscopy as a function of $V_{\mathrm{reso}}$, and $V_{\mathrm{G4}}$, respectively. The observed features correspond to the qubit frequency $f_q$. The dashed curves in Figs.~\ref{fig1}E and F show fits to the qubit frequency as a function of $V_{\mathrm{reso}}$ and $V_{\mathrm{G4}}$ respectively, following Eqs.~\ref{eq:E}--\ref{eq:alpha_DC}, yielding $\alpha_{\mathrm{reso}}^\mathrm{DC} = (1.345 \pm 0.006) \times 10^{-2}$ and $\alpha_{\mathrm{G4}}^\mathrm{DC}=(3.64 \pm 0.02) \times 10^{-3}$. Electrodes G1--G3 show negligible coupling to the qubit. 
	
	The parameters $\alpha_{\mathrm{reso}}^\mathrm{AC}$, $\alpha_{\mathrm{reso}}^\mathrm{DC}$, and $\alpha_{\mathrm{G4}}^\mathrm{DC}$ depend on the electron position, allowing us to infer the trapping location, albeit with multiple possible solutions. The inferred electron positions span $x = 3.4$--$5.5$~nm and $y = 0.02$--$0.57$~nm, with tilt angles $\theta \leq 15^\circ$ and potential minima separations $d = 50$ and $100$~nm, covering the plausible parameter range (see Table~\ref{tab:e-place-F} in the Supplementary Materials for more details). The mean neon thickness is estimated to be $1.3$~nm from the resonator frequency shift (see Supplementary Materials~\ref{sec:position_estimation}), and the electron floats $1$--$2$~nm above the neon surface~\cite{Zhou2022-nk,Kanai2024-bo}, giving $z \approx 2$--$3$~nm; although the local thickness at the electron position is unknown due to surface roughness, the inferred $x$ and $y$ positions are weakly dependent on $z$ in the range of $1$--$10$~nm. The light-blue sphere in Fig.~\ref{fig1}B indicates the representative electron position $(x,y)=(5.37,0.57)$~nm, which is used later for the spin qubit fidelity estimation in Sec.~\ref{sec:Prospect_spin}.

	We note that, due to charge instability in the device environment, the gate voltage required to tune the charge qubit frequency into resonance with the resonator can change over time. In practice, the operating point typically remains stable for about a day, after which a discrete shift in the required gate voltage is sometimes observed. Importantly, these shifts do not noticeably affect the coupling strengths between the electron and the electrodes or the resonator. This suggests that the electron position relative to the device electrodes remains approximately unchanged. One possible origin of the observed switching behavior is slow rearrangements of nearby charges in the device environment, which modify the local electrostatic potential without significantly altering the electron’s spatial confinement.

	\subsection{Coherent control of the charge qubit}
\begin{figure}[t]
		\centering
		\includegraphics[width=\columnwidth]{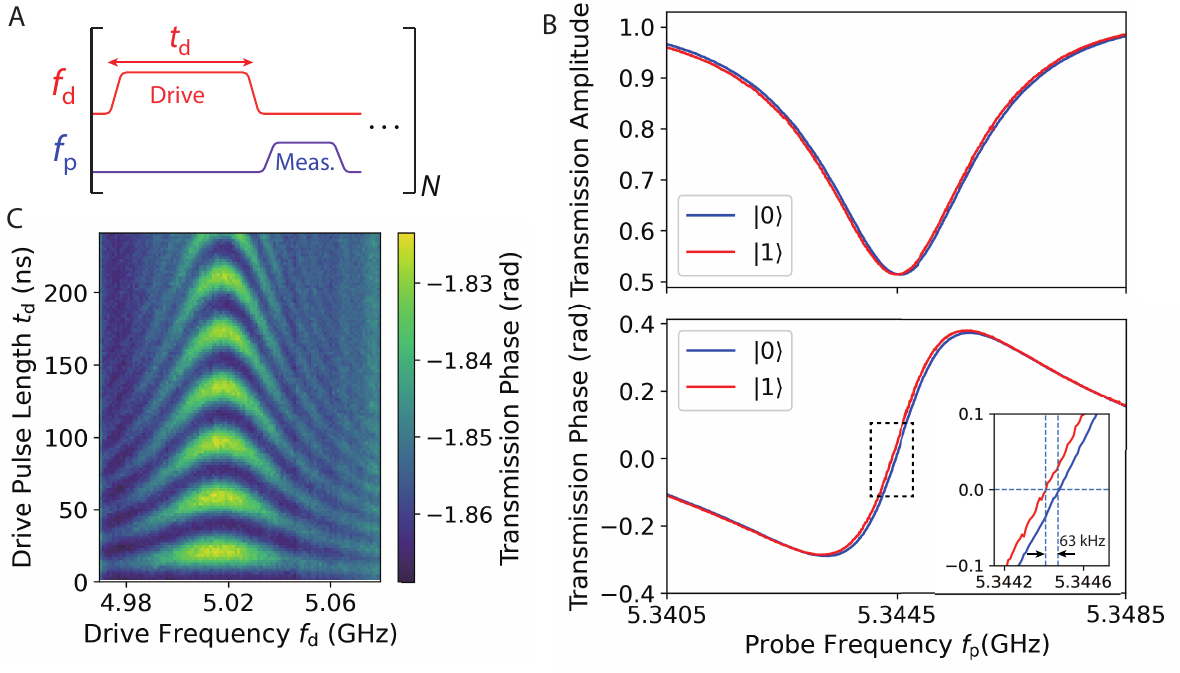}
		\caption{(A) Schematic of the Rabi measurement sequence. A drive pulse at frequency $f_d$ with length $t_d$ is followed by a probe pulse at frequency $f_p$ with a length of 2~$\mu$s. Each data point is averaged over $N=\mathrm{300}$ measurements with a waiting time of 1~ms between repetitions. The drive strength is set to $A/A_0 = 0.3$.  (B) Transmission amplitude and calibrated phase of the probe signal as a function of the probe frequency $f_p$ around $f_r$ for the electron in states $|0\rangle$ (blue, no drive pulse) and $|1\rangle$ (red, with a $\pi$ drive pulse), showing the dispersive shift of the resonator frequency of 63~kHz depending on the qubit state. (C) Rabi oscillations of the qubit measured by sweeping the drive frequency $f_d$ around $f_q$, showing the characteristic chevron pattern. The probe frequency is set to $f_p=f_r$. The Rabi frequency is $f_{\mathrm{Rabi}} = 25.9\,\mathrm{MHz}$, and the Rabi decay time, defined as the $1/e$ decay time constant of the Rabi oscillation envelope, is $T_2^{\mathrm{Rabi}} = 1.1\,\mu\mathrm{s}$.}
		\label{fig2}
	\end{figure}

	Here we demonstrate qubit control at $\varepsilon = 0$, the charge sweet spot where the qubit coherence is optimal~\cite{Zhou2023-iw}, using the ground and first excited orbital states as $|0\rangle$ and $|1\rangle$ (Fig.~\ref{fig1}C). The qubit Hamiltonian is reduced to $H = (\hbar\Delta/2)\sigma_x$, and the qubit frequency is $f_q = \Delta/2\pi = 5.02\,\mathrm{GHz}$. To control the qubit state and subsequently measure it, we first apply a drive pulse and then perform the readout using a probe pulse (Fig.~\ref{fig2}A). Due to the dispersive coupling between the qubit and the resonator, the resonator frequency shifts by 63~kHz depending on whether the qubit is in the $|0\rangle$ or $|1\rangle$ state, corresponding to a phase shift of approximately 0.035~rad in the transmitted signal after phase calibration (Fig.~\ref{fig2}B). As a result, the qubit state can be measured from the phase change of the transmitted signal at $f_r$. To demonstrate Rabi oscillations~\cite{Rabi}, we vary the drive pulse length (Fig.~\ref{fig2}C). The qubit then undergoes coherent oscillations between $|0\rangle$ and $|1\rangle$. The drive power at room temperature is set to 10~dBm for all pulsed measurements in this manuscript. The room-temperature attenuator sets the drive amplitude to $A/A_0$, where $A/A_0$ is the voltage ratio relative to the full drive amplitude at 10~dBm.

\begin{figure}[h]
		\centering
		\includegraphics[width=\linewidth]{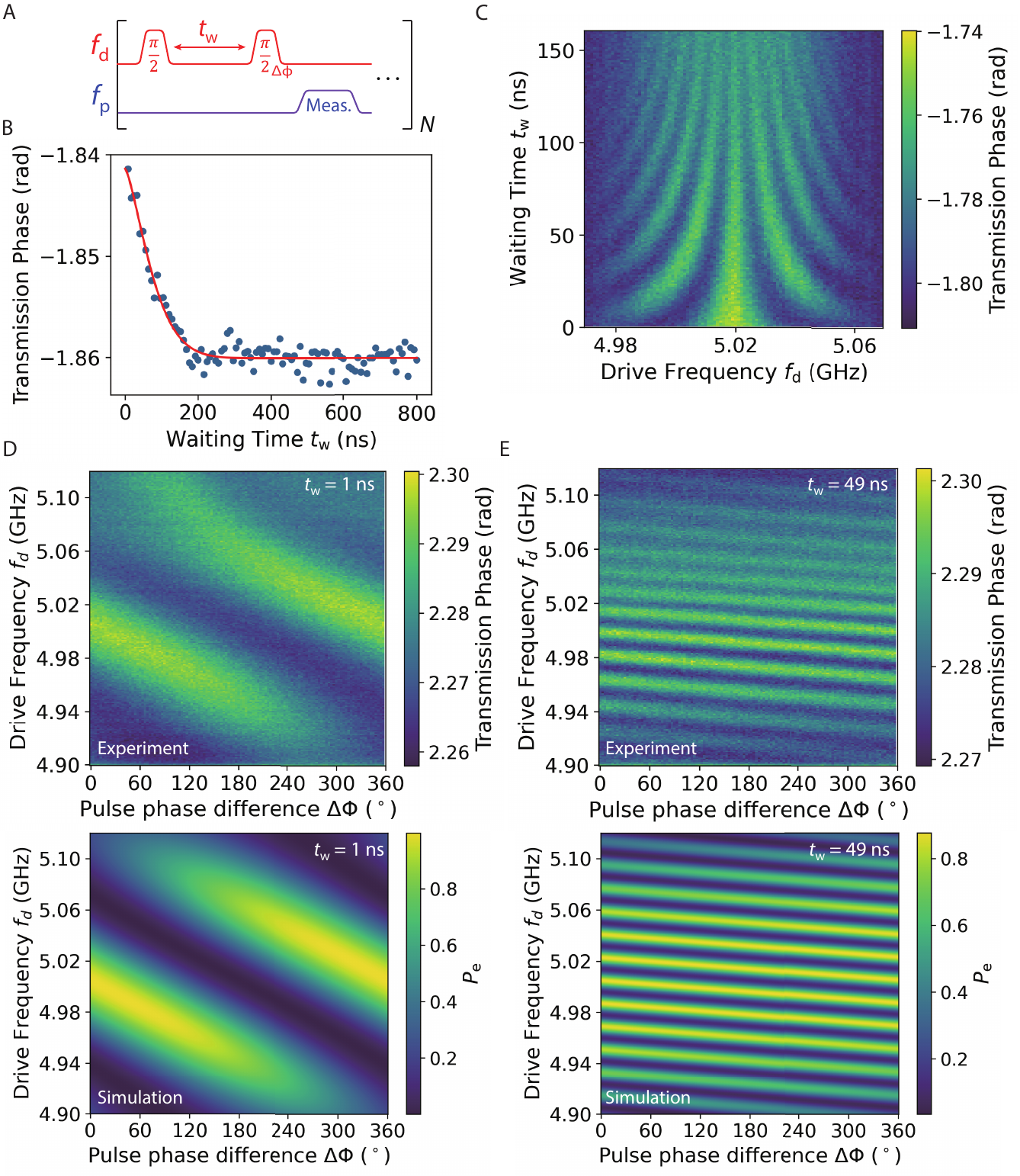}
		\caption{(A) Schematic of the Ramsey measurement sequence. Two $\pi/2$ pulses are applied with a waiting time $t_\mathrm{w}$ and a relative phase difference $\Delta\Phi$ between them. A probe pulse with a pulse duration of 1~$\mu$s is applied a few hundred nanoseconds after the second $\pi/2$ pulse. Each data point is averaged over $N=100\sim1000$ measurements with a waiting time of 1~ms between repetitions.  (B) Ramsey decay at $\Delta\Phi = 0$. The solid line shows a fit to $\exp\left[-(t_\mathrm{w}/T_2^*)^\beta \right]$, yielding $\beta = 1.4 \pm 0.4$ and $T_2^* = 91\pm 23$~ns. (C) Ramsey fringes measured by varying the drive frequency $f_d$ and the waiting time $t_\mathrm{w}$ between the two $\pi/2$ pulses (D,E) Demonstration of $\pi/2$ rotations around an arbitrary axis in the $xy$ plane of the Bloch sphere by changing the relative phase difference $\Delta\Phi$, with $t_\mathrm{w}=1$~ns in (D) and $t_\mathrm{w}=49$~ns in (E). The experimental results are well reproduced by the simulation. The Rabi frequency is $f_\mathrm{Rabi}=14.286$~MHz in (B, C) and 45.45~MHz in (D, E). The $\pi/2$ pulse length is 17.5 ns in (B, C) and 5.5 ns in (D, E).
			\label{fig3}}
	\end{figure}

	Next, to demonstrate two-axis control and characterize the qubit coherence, we perform Ramsey interferometry~\cite{Ramsey}. Two $\pi/2$ pulses are applied with a waiting time $t_\mathrm{w}$ between them and a relative phase difference $\Delta\Phi$. The qubit state is then read out using a probe pulse (Fig.~\ref{fig3}A). From a standard Ramsey decay measurement, we extract $T_2^* = 91\pm 23$~ns (Fig.~\ref{fig3}B), which is significantly shorter than previously reported values~\cite{Zhou2022-nk,Zhou2023-iw}; possible reasons are discussed in Sec.~\ref{sec:T2_T1} and Sec.~\ref{sec:discussion}. When the drive frequency $f_d$ is swept around $f_q$, Ramsey fringes appear (Fig.~\ref{fig3}C). By varying the relative phase $\Delta\Phi$ between the two $\pi/2$ pulses, the rotation axis of the second pulse can be tuned within the $xy$ plane of the Bloch sphere, enabling two-axis control of the qubit (Fig.~\ref{fig3}D, E). The fringes observed along the drive frequency $f_d$ axis are governed by the phase accumulation arising from the detuning between $f_d$ and $f_q$ over the entire pulse sequence. While a simple estimate would suggest a fringe period of approximately $1/t_\mathrm{w}$ based on free evolution during the waiting time, this approximation is not valid in our parameter regime. In particular, the condition $|f_d - f_q| \ll f_\mathrm{Rabi}$ is not satisfied, and the full driven dynamics, including the finite-duration pulses, must be taken into account. An analytical model incorporating the full time evolution during both the pulses and the waiting time quantitatively reproduces the experimental data (Fig.~\ref{fig3}D,E).

	\subsection{Nonlinear Qubit Frequency Shift under Strong Driving}
	
\begin{figure}[h!]
		\centering
		\includegraphics[width=\linewidth]{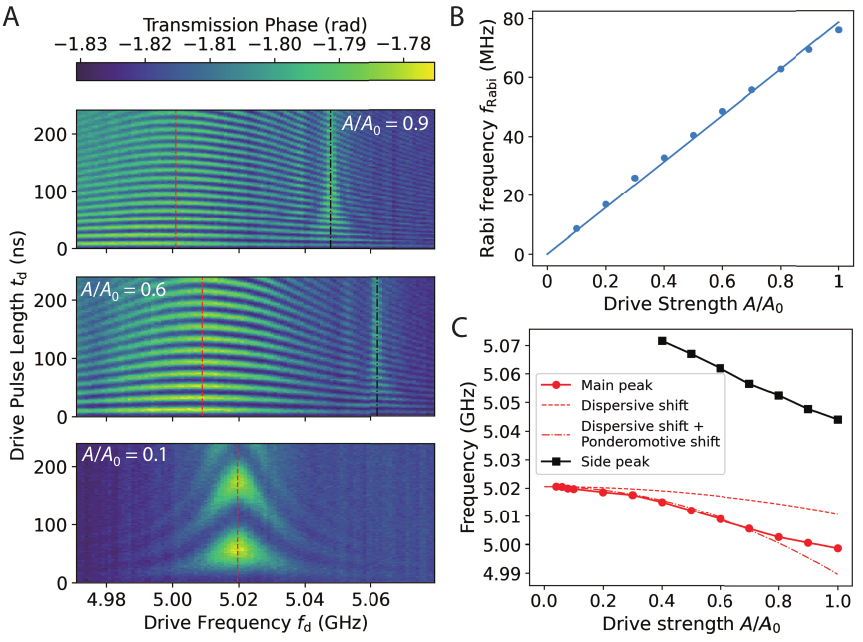}
		\caption{ Strong drive regime. The pulse sequence used here is the same as that shown in Fig.~\ref{fig2}A. (A) Rabi oscillations for varying drive strength $A/A_0$. The Rabi-chevron pattern of the resonance (referred to as the main peak) is clearly visible. With increasing drive strength, this resonance shifts toward lower frequencies, indicating a reduction of the qubit frequency $f_q$. A side peak also emerges on the higher-frequency side. Vertical dashed red and black lines indicate the center frequencies of the main and side peaks, respectively. (B) Rabi frequency of the main peak as a function of the drive strength; the solid line shows a linear fit. The uncertainty in the Rabi frequency extracted from each Rabi oscillation is smaller than the size of the data points. (C) Experimentally observed resonance-frequency shifts of the main peak (filled red circles) and side peak (filled black rectangles) connected by solid lines, as a function of the drive strength $A/A_0$. The dashed line shows the dispersive shift induced by photons in the resonator, while the dash-dotted line represents the combined ponderomotive and dispersive shift.
			\label{fig4}}
	\end{figure}

	At higher drive powers, the system response deviates from the simple linear behavior observed at low power, revealing additional features in the resonance spectrum. As shown in Fig.~\ref{fig4}A, when a higher drive power is applied, the Rabi frequency $f_{\mathrm{Rabi}}$ increases linearly with the drive strength, as expected (Fig.~\ref{fig4}B), reaching $76~\mathrm{MHz}$ at $A/A_0 = 1$. In addition, anomalous features emerge: the qubit frequency $f_q$ shifts toward lower frequencies as the drive strength increases (Fig.~\ref{fig4}C). We consider possible origins of this frequency shift. One plausible explanation is a dispersive shift caused by drive photons remaining in the resonator. The expected shift is given by (Supplementary Materials~\ref{sec:dispersive})
	\begin{equation}
		\Delta f_q = \frac{f_\mathrm{Rabi}^2}{2(f_q-f_r)},
	\end{equation}
	where $f_q - f_r=-324$~MHz is the detuning between the qubit and resonator frequencies. Note that the conventional dispersive shift~\cite{Schuster2005-le} is typically induced by the weak probe tone used for qubit readout. In contrast, here we consider the population of resonator photons generated by the strong drive pulse itself, which can also produce an AC Stark (dispersive) shift of the qubit frequency. The estimated frequency shift is shown as the dashed line in Fig.~\ref{fig4}C. Although its magnitude is comparable to the experimentally observed shift at the order-of-magnitude level, the calculated dispersive shift is smaller than the  measured value, indicating that the dispersive shift is present but does not fully account for the observation. The residual discrepancy suggests that an additional mechanism is at play. One candidate is a ponderomotive modification of the trapping potential caused by the spatially inhomogeneous microwave electric field \cite{Kibble1966-ph,cook1985pra,Bucksbaum1987-pd}. This mechanism would require the electron to be located closer to the edge of the resonator than the position shown in Figs.~\ref{fig1}B and C (see Supplementary Materials~\ref{sec:Ponderomotive}). In that case, the AC electric field amplitude would have a maximum near the center of the double-well potential, leading to a ponderomotive reduction of the effective confinement strength and a resulting downward shift of the qubit transition frequency. The estimated frequency shift resulting from this mechanism, combined with the dispersive shift discussed above, is shown as the dash-dotted line in Fig.~\ref{fig4}C (see Supplementary Materials~\ref{sec:Ponderomotive} for more details). While the ponderomotive effect may be a possible contributor to the residual shift, its role remains inconclusive and warrants further investigation.

	Furthermore, at higher drive powers an additional side peak appears at a slightly higher frequency than the main peak (Fig.~\ref{fig4}A). Its resonance frequency also shifts toward lower frequencies with increasing drive power, with a stronger power dependence than that of the main peak (Fig.~\ref{fig4}C). We initially considered whether this feature corresponds to the $|1\rangle \rightarrow |2\rangle$ transition, where $|2\rangle$ is the second excited orbital state. To test this possibility, we prepared the electron in the first excited state using a $\pi$ pulse at the main peak frequency and then searched for the corresponding transition. However, no clear signal associated with such a transition was observed. This suggests that the side peak is unlikely to originate from the $|1\rangle \rightarrow |2\rangle$ transition. Instead, it may arise from another nearby electron with a similar qubit frequency. If so, the electron may be trapped at a different location and experience a different local confinement potential, causing its resonance frequency to shift with drive power in a different way and qualitatively explaining the observed behavior. Nevertheless, the precise origin of this side peak remains unclear.

	\subsection{Relaxation and coherence\label{sec:T2_T1}}
	
\begin{figure}[ht!]
		\centering
		\includegraphics[width=\linewidth]{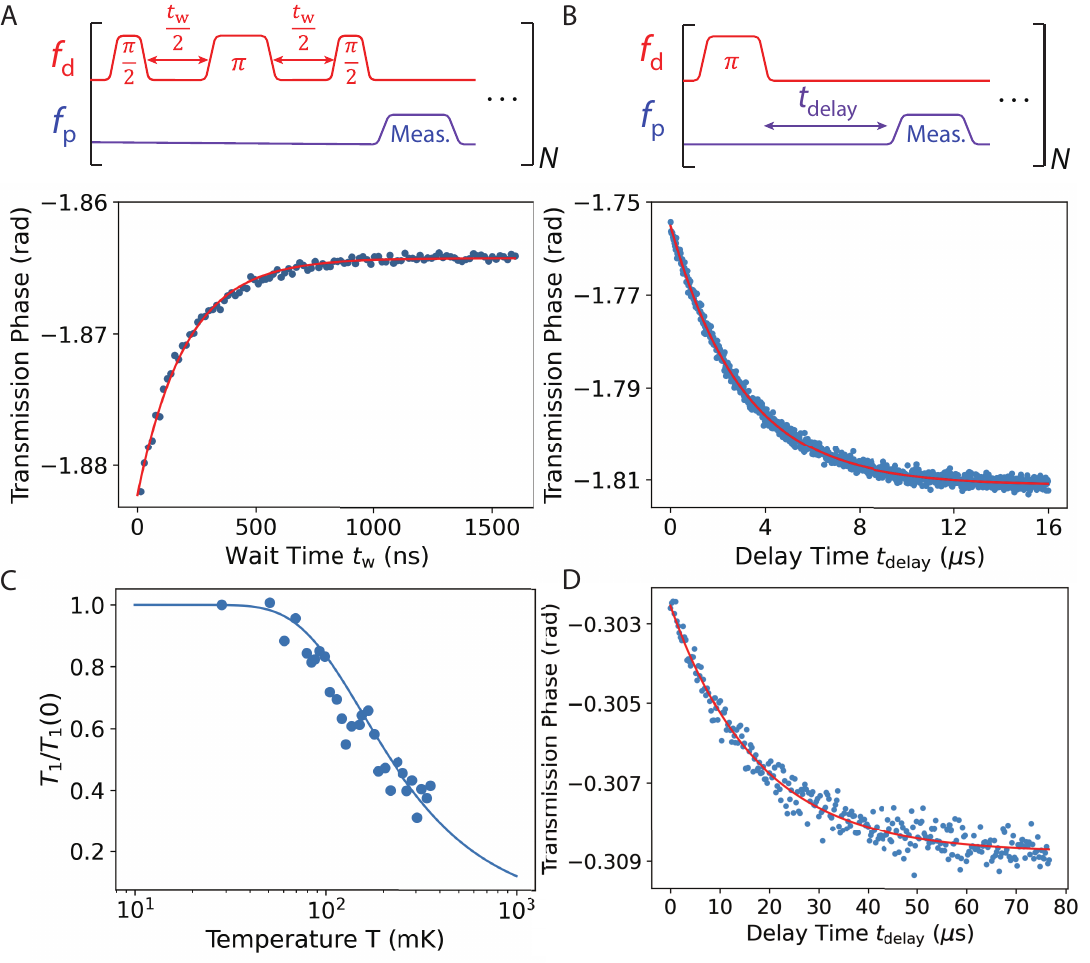}
		\caption{(A) Hahn-echo measurement. The decay is fitted with a single-exponential form $\exp(-t/T_2)$, yielding $T_2 = 202\pm5$~ns. (B) Energy relaxation measurement. The solid line is a fit to $\exp(-t_\mathrm{delay}/T_1)$, yielding $T_1 = 3.056 \pm 0.015~\mu\text{s}$. 
			(C) Temperature dependence of the relaxation time $T_1$, fitted by $T_1(T)=T_1(0)\tanh[hf_q/(2k_\mathrm{B}T)]$. 
			(D) Energy relaxation measurement after annealing the solid neon by warming it to 8.6~K, cooling back to below 10~mK, and redepositing electrons. The measured qubit here corresponds to a different single-electron realization from those used elsewhere in this work. The solid line is a fit to $\exp(-t_\mathrm{delay}/T_1)$, yielding $T_1 = 17.7\pm 0.6~\mu\text{s}$.
		}\label{fig:coherence_time}
	\end{figure}

	Finally, we characterize the relaxation and coherence properties of the charge qubit using pulse sequences (Fig.~\ref{fig:coherence_time}). 
	To extend the dephasing time $T_2^* = 91$~ns measured in Fig.~\ref{fig3}B, we performed a Hahn-echo measurement~\cite{Hahn1950}. 
	However, the coherence time increases only slightly to $T_2 = 202$~ns, and the decay follows a single-exponential form (Fig.~\ref{fig:coherence_time}A). 
	This indicates that inserting additional $\pi$ pulses would not significantly improve the coherence time, suggesting that the noise is dominated by largely uncorrelated (white) noise rather than slow fluctuations. 
	Both $T_2^*$ and $T_2$ are significantly shorter than previously reported values for electrons on solid neon~\cite{Zhou2023-iw,Li2025-ci,Li2025-em}. One possible explanation is the presence of additional trapped electrons in the vicinity of the qubit, which can generate charge noise. 
	Indeed, the electron responsible for the side peak observed in Fig.~\ref{fig4} has a qubit frequency close to that of the main qubit and may contribute to such fluctuations.
	
	We measure the energy relaxation time $T_1$ by preparing $|1\rangle$ with a $\pi$ pulse and monitoring the return to $|0\rangle$ as a function of delay time $t_\mathrm{delay}$. 
	The extracted $T_1$ is $3.06~\mu$s (Fig.~\ref{fig:coherence_time}B). 
	Following Ref.~\onlinecite{Li2025-em}, we also measure the temperature dependence of $T_1$ and observe a similar behavior, which is consistent with relaxation dominated by phonon emission (Fig.~\ref{fig:coherence_time}C). After these measurements, we warmed the system to 8.6~K to anneal the solid neon, cooled it back below 10~mK, and redeposited electrons. Repeating the relaxation measurement focusing on one of the newly deposited electrons as the qubit yields a longer relaxation time of $T_1 = 17.7~\mu\text{s}$. At the same time, the coupling parameters changed from 
	$\alpha_{\mathrm{reso}}^\mathrm{DC} = 1.35 \times 10^{-2}$ and $\alpha_{\mathrm{G4}}^\mathrm{DC} = 3.6 \times 10^{-3}$ before annealing to $\alpha_{\mathrm{reso}}^\mathrm{DC} = 1.88 \times 10^{-3}$ and $\alpha_{\mathrm{G4}}^\mathrm{DC} = 9.8 \times 10^{-4}$ after annealing. For comparison, Ref.~\onlinecite{Li2025-em} reports $T_1 \sim 10~\mu$s with $\alpha \sim  4.1\times10^{-4}$, while Ref.~\onlinecite{Zhou2023-iw} reports $T_1 \sim 100~\mu$s with $\alpha \sim  1.7\times10^{-6}$. These results suggest that the relaxation time may be correlated with the coupling strength between the electron and the electrodes. Combined with the observed temperature dependence, which indicates phonon-dominated relaxation, this suggests that energy relaxation may occur via coupling of the electron to phonons in the electrodes and/or the Si substrate. One possible explanation for the observed increase in $T_1$ after annealing is an increase in the neon film thickness, which may increase the distance between the electron and the electrodes and/or the Si substrate, thereby reducing the coupling strength $\alpha$ and leading to a longer $T_1$. Note that, at present, annealing should be viewed as a way to modify the overall environment, rather than a deterministic method for improving coherence, as its impact on coherence has not yet been systematically studied. Nevertheless, these observations suggest that controlling the neon film thickness could provide a way to systematically tune the trade-off between relaxation and electrode coupling. Achieving such control will likely require a more controlled neon deposition technique, as the present deposition through the liquid phase does not allow precise control of the film thickness~\cite{Duthaluru2025-ym}.

	\subsection{Prospects for Spin-Qubit Realization \label{sec:Prospect_spin}}
	
\begin{figure}[ht!]
		\centering
		\includegraphics[width=\linewidth]{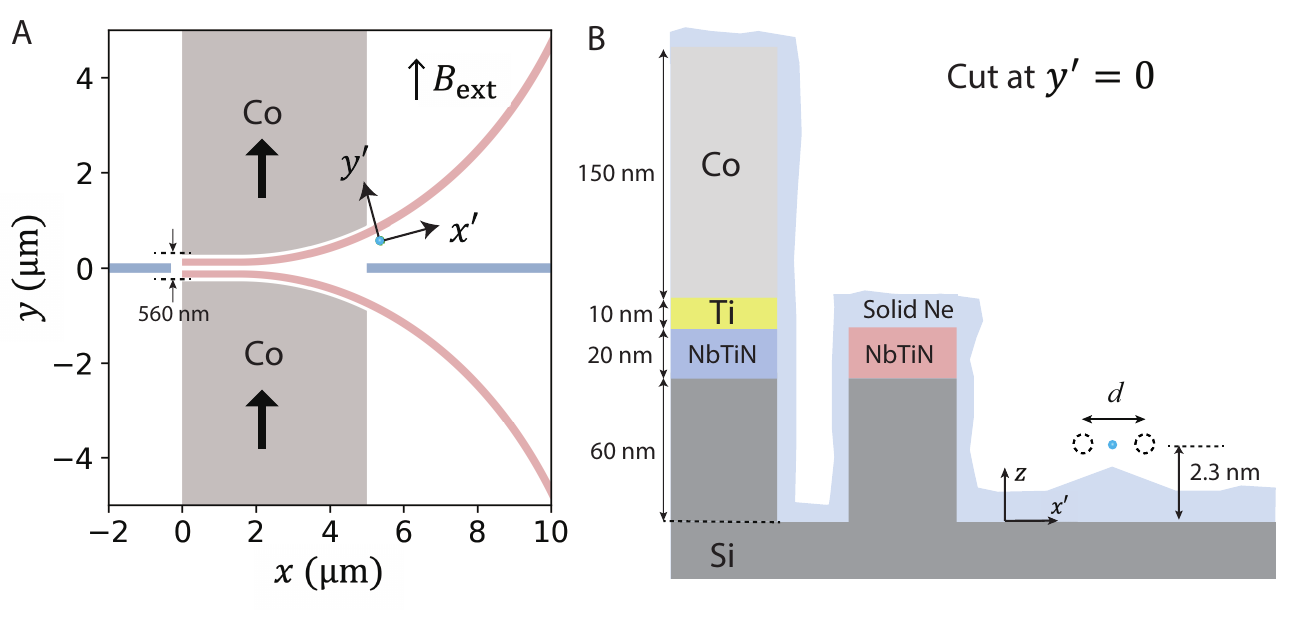}
		\caption{(A) Schematic of the device geometry including the ferromagnets integrated near the electron confinement region. The electron position is $(x,y)=(5.37,0.57)$~nm, with potential minima separation $d=100$~nm and tilt angle $\theta=15^\circ$. (B) Cross-section of the device and potential along $y'=0$. The Co ferromagnet (150~nm) with a Ti adhesion layer (10~nm) is integrated near the electron confinement region. The electron--Si substrate distance is 2.3~nm, corresponding to a neon thickness of 1.3~nm.
		}\label{fig:spin}
	\end{figure}

	Based on the experimentally determined electron position, we now consider the expected performance of a spin qubit realized by incorporating ferromagnets into the device. We focus on the solution with $(x, y) = (5.37, 0.57)$~nm, $d =100$~nm, and $\theta = 15^\circ$, assuming a neon thickness of 1.3~nm (see Supplementary Materials~\ref{sec:position_estimation}) and an electron--neon surface distance of 1~nm following Ref.~\onlinecite{Kanai2024-bo}, giving an electron--Si substrate distance of $z=2.3$~nm (Fig.~\ref{fig:spin}A, B). Although the neon thickness at the electron position is not precisely known experimentally, we find that the estimated spin Rabi frequency and spin-qubit fidelity discussed below remain nearly unchanged even when assuming, for example, a larger electron--Si distance of 7 nm. This weak dependence arises because the relevant magnetic-field gradients vary only moderately over this distance range.
	
	The ferromagnet geometry, shown in Fig.~\ref{fig:spin}A, B, follows Ref.~\onlinecite{Tian2025-fz}, where the optimized stack for the electron configuration considered here consists of Ti (10~nm)/Co (150~nm). Here, we theoretically examine electric dipole spin resonance (EDSR) at $\varepsilon=0$ \cite{Tokura2006,Pioro-Ladriere2007,Pioro-Ladriere2008,Benito2019-pi,Croot2020-mv} driven by microwaves delivered through the feedline to the resonator, analogous to the charge-qubit driving scheme implemented in this work, without relying on an independent drive gate, in contrast to Refs.~\onlinecite{Tian2025-fz,Samkharadze2018,Petersson2012,Dijkema2025}. An external magnetic field $B_{\mathrm{ext}}$ applied along the $y$ direction magnetizes the Co ferromagnets along the same direction, producing a stray magnetic field at the electron position. We consider EDSR at $B_{\mathrm{ext}} = 0$ after sweeping to a sufficiently large value, assuming for simplicity that the Co layer remains fully magnetized, neglecting any reduction in remanent magnetization. At this point, two components of the stray magnetic field become relevant. Expressed in frequency units through the electron gyromagnetic ratio, the average field component experienced by the electron along the $y$-axis is estimated to be $b_\parallel/2\pi = 417~\mathrm{MHz}$, whereas the difference in the perpendicular component between the two potential minima is estimated to be $b_\perp/2\pi = 168~\mathrm{MHz}$. The spin Rabi frequency is given by $f_{\mathrm{Rabi,spin}} = \Lambda f_{\mathrm{Rabi}}$, where $\Lambda = \sin\bar{\phi}$ with $\bar{\phi}$ being the spin-charge mixing angle~\cite{Benito2019-pi},
	\begin{align}
		\bar{\phi} = (\phi_+ + \phi_-)/2, \, \, \, \, \, \, 
		\phi_\pm = \arctan\frac{b_\perp}{\Delta \pm b_\parallel}.
	\end{align}
	For the above values of $b_\parallel$ and $b_\perp$, we obtain $\Lambda=0.0328$. With $f_{\mathrm{Rabi}}=76~\mathrm{MHz}$, the corresponding spin Rabi frequency is estimated to be $f_{\mathrm{Rabi,spin}}=2.49~\mathrm{MHz}$.
	
	Furthermore, we evaluate the average gate fidelity of a $\pi$ pulse over all possible input states as a representative example of single-qubit gate fidelity~\cite{Benito2019-pi,Tian2025-fz}:
	\begin{equation}
		\bar{F_1} = \frac{1}{6}
		\left[
		3 + e^{-2 t_g \gamma_s'}
		+ 2 e^{- t_g \gamma_s'} e^{-\frac{(t_g{\gamma_s^*}')^2}{2}}
		\right], \label{eq:single_qubit_fidelity}
	\end{equation}
	where $\gamma_s'=\Lambda^2 \gamma_c+ (1-\Lambda^2) \gamma_s$, $t_g = \frac{1}{2 f{\mathrm{Rabi,spin}}}$, and
	\begin{align}
		{\gamma_s^*}'=\sqrt{\left(\frac{\Lambda^2{\gamma_c^*}^2}{b_\parallel}\right)^2 
			+\left(\frac{\cos\phi_++\cos\phi_-}{2}\gamma_s^*\right)^2}
	\end{align} 
	is the quasistatic spin loss rate~\cite{Benito2019-pi, Tian2025-fz}. Here, we use $\gamma_c/2\pi = 1/(2T_1)$, where $T_1 = 3.056~\mu\mathrm{s}$ is the charge relaxation time measured in this work,  $\gamma_c^*=1/T_2^*$, where $T_2^*=91$~ns is the charge dephasing time measured in this work. For the spin parameters, we use the spin relaxation rate $\gamma_s/2\pi = 5.88~\mathrm{kHz}$ and spin dephasing rate $\gamma_s^*=6.25$~kHz, corresponding to the largest values reported in Refs.~\onlinecite{Chen2022-on,Tian2025-fz}, which were attributed to thermal magnetic noise in copper. This yields a conservative fidelity estimate of $\bar{F}_1 = 99.52\%$.
	
	The tunability of $\Lambda$ via the external magnetic field $B_{\mathrm{ext}}$, as well as other possible electron trapping positions and configurations with slightly different neon thicknesses arising from surface roughness or nonuniformity, are discussed in Supplementary Materials~\ref{sec:EDSR}. We find that the estimated maximal fidelity $\bar{F}_\text{1,max}=99.83\%$ by changing $B_{\mathrm{ext}}$ remains largely unchanged across these configurations. Please note that the lower fidelity compared to Ref.~\onlinecite{Tian2025-fz} originates from the faster charge $T_1$ relaxation in our device. If we instead use the improved post-annealing value of $T_1 = 17.7~\mu\mathrm{s}$ and the spin relaxation rate $\gamma_s/2\pi = 33.3~\mathrm{Hz}$, an intermediate value reported in Refs.~\onlinecite{Chen2022-on,Tian2025-fz}, which was attributed to hyperfine interactions with nuclear spins in natural neon, the fidelities become $\bar{F}_{\mathrm{1,max}} = \mathrm{99.995}\%$.

	\section{Discussion \label{sec:discussion}}
	
	We demonstrated coherent control of a single-electron charge qubit floating in vacuum on the surface of solid neon using a NbTiN nanowire resonator. At high drive power, we observe a Rabi frequency of up to 76~MHz and a downward shift in the qubit frequency, which we presumably attribute to the dispersive shift due to the high number of photons in the resonator and to the ponderomotive effect.
	
	The electron is trapped in a potential likely dominated by microscopic neon surface structures and/or charges in the neon or substrate, while being partially controlled by DC voltages applied to the electrodes. Electrostatic disorder and uncertainty arising from the former complicate deterministic positioning of the electron and pose a challenge for scalability. To partially address this issue, we used the position-dependent coupling strength to the electrodes as a diagnostic tool to infer the electron position. Further improvements in the growth of the neon thin film will be required to achieve better control of the trapping potential.
	
	The measured coherence times $T_2^*$ and $T_2$ are significantly shorter than previously reported values for electrons on solid~\cite{Zhou2023-iw,Li2025-ci,Li2025-em}. The Hahn-echo measurement provides only a modest improvement, suggesting that the dephasing is dominated by broadband (approximately white) noise rather than slow fluctuations. One possible origin of such noise is the presence of additional stray electrons in the vicinity of the qubit, which may generate charge fluctuations over a wide range of frequencies. Improved control of the electron environment will therefore be important. For example, in the present device the Si substrate is partially exposed; placing additional electrodes over these regions and biasing them with negative voltages could help prevent unwanted electrons from entering the trapping region.
	
	The observed temperature dependence of $T_1$ suggests coupling to thermal phonons in the environment. This indicates that the relaxation rate may be determined by how strongly the electron couples to phonons in the electrodes or substrate. Achieving systematic control of the neon film thickness will be important for clarifying and mitigating this relaxation mechanism.
	
	Overall, these results suggest that improved control of the qubit environment, including the neon thin film and the surrounding electron population, will be essential for enhancing the performance of electron-on-neon qubits. At the same time, the realization of spin qubits in this platform, which are expected to provide substantially longer coherence times than charge qubits, remains an important future direction~\cite{Tian2025-fz,Chen2022-on}. Our estimates indicate that spin-qubit demonstrations remain feasible even without deterministic electron trapping. Pursuing these two directions in tandem represents a promising path toward high-performance electron-on-neon qubits.

	\section{Materials and Methods}
	\subsection{Validity of the Double-Well Approximation \label{sec:double_well_potential}}
	
\begin{figure}[ht!]
		\centering
		\includegraphics[width=\columnwidth]{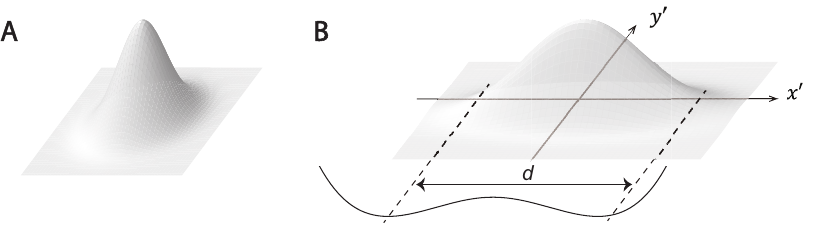}
		\caption{(A) Schematic illustration of an isotropic bump on the solid neon surface proposed in Ref.~\onlinecite{Kanai2024-bo}. (B) Schematic illustration of an elliptical bump. The solid line illustrates the double-well potential along the $x'$ direction, and the distance between the two potential minima is denoted by $d$. In Fig.~\ref{fig1}C, the dashed circles indicate the positions $(x',y')=(\pm d/2, 0)$. These figures are intended as qualitative schematics and are not quantitatively accurate. \label{fig:Elliptical_bump}}
	\end{figure}

	Ref.~\onlinecite{Kanai2024-bo} proposed that electrons trapped on solid neon are localized by nanoscale isotropic bumps on the neon surface as shown in Fig.~\ref{fig:Elliptical_bump}(A). In a more general situation, however, the bump is expected to be elliptical rather than perfectly isotropic. 
	For an elliptical bump (Fig.~\ref{fig:Elliptical_bump}(B)), the confinement along the short axis ($y'$) becomes stronger than that along the long axis. 
	We assume that the qubit is formed by the weaker confinement along the long axis ($x'$), and model the potential along this direction as a double-well potential. In addition, we assume that the confinement potential can be approximated as separable in the $x'$ and $y'$ directions.
	
	Note that, although we have assumed that electrons trapped on solid neon are localized by nanoscale isotropic bumps on the neon surface, we do not exclude other possible microscopic origins. For example, a similar double-well landscape could also arise from two nearby positive charges in the substrate or within the neon.
	
	\subsection{Device fabrication}
	
\begin{figure}[!ht]
		\centering
		\includegraphics[width=\columnwidth]{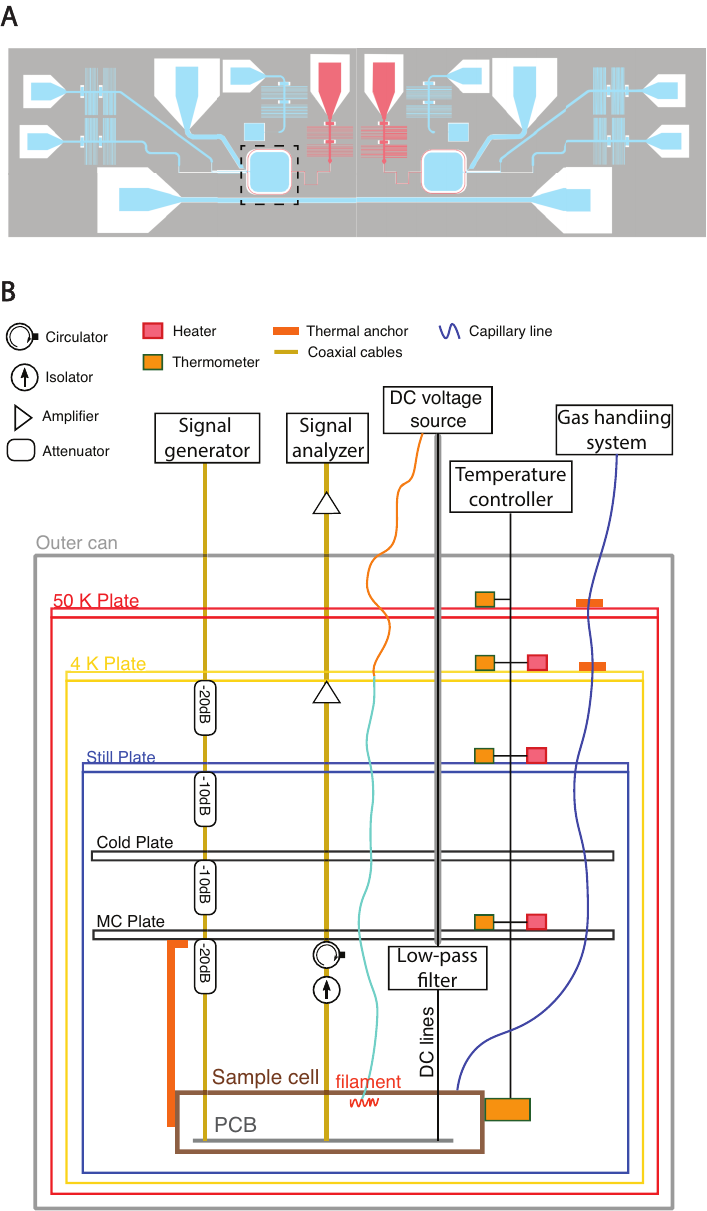}
		\caption{(A) Device layout. The resonator center conductor is shown in pink, 
			the ground plane in gray, and the other electrodes in light blue. 
			The dashed rectangle indicates the region corresponding to Fig.~\ref{fig1}A; 
			the resonator inside this region is used in this work. (B) Schematic of the wiring in the dilution refrigerator. The high-frequency input line is 
			step-attenuated at each temperature stage (total attenuation $\sim -60$ dB). 
			The output signal passes through a circulator--isolator pair, a 4 K HEMT 
			amplifier, and a room-temperature amplifier. DC control signals are delivered 
			to the cell through a multichannel low-pass filter circuit consisting of 
			$\pi$-filters and RC filters connected in series at the mixing-chamber plate. 
			An additional DC source powers the temperature-control unit and the tungsten filament; the wiring transitions from copper (orange) to superconducting wire (light blue) below the 4~K plate. The stainless-steel capillary filling line (blue) connects the cell 
			to a homemade neon gas-handling system.
		}\label{fig:Bluefors_setup}
	\end{figure}

	The superconducting nanowire resonator and DC electrodes were fabricated 
	from a 20-nm-thick NbTiN film deposited on a high-resistivity silicon substrate 
	and patterned by reactive-ion etching (RIE). The detailed nanofabrication 
	process is described in Ref.~\onlinecite{Tian2025-fz}. Compared with Ref.~\onlinecite{Tian2025-fz}, additional electrodes were introduced to apply DC voltages. All DC electrodes except G3 are equipped with on-chip filters~\cite{Mi2017-go} (see Supplementary Materials ~\ref{sec:dc_filter_transmission} for filter characterization). Figure~\ref{fig:Bluefors_setup}A shows the overall device layout. 
	The chip contains two resonators, and one of them is used in this work. After fabrication, the chip was glued onto a copper printed circuit board (PCB). Aluminum wire bonds connect the resonator feedline and DC electrodes to the PCB, which is sealed inside a copper sample cell (Fig~\ref{fig:Bluefors_setup}B).
	
	\subsection{Measurement setup}
	
	The device was measured in a dilution refrigerator (Bluefors LD400) with a base temperature below 10~mK. The microwave input line has a total of 77.7~dB of attenuation, including the line attenuation itself measured at room temperature. At the base temperature of 10~mK, we have Eccosorb filters and RLC Electronics low-pass filters (F-30-10-R). The output line contains a cryogenic circulator chain and a cryogenic amplifier at 4~K, followed by a room-temperature amplifier. DC electrode voltages were supplied by a QDevil ultra-low-noise DAC through low-pass filters placed at the MC plate and low-pass filters placed on the chip to reduce high-frequency noise.
	
	The resonator transmission $S_{21}$ was measured using either a Keysight vector network analyzer or a Zurich Instruments SHFQC qubit controller. For the two-tone measurements, the drive and probe signals were combined using a power combiner at room temperature and sent through the same input line. For pulsed measurements, square-envelope microwave pulses were generated and mixed internally using the arbitrary waveform generator of the SHFQC.

	\subsection{Solid neon deposition and electron loading}
	
	Solid neon was deposited onto the chip by introducing high-purity neon gas 
	through a capillary line (Fig.~\ref{fig:Bluefors_setup}B) while the sample 
	cell temperature and pressure were maintained on the liquid side of the 
	neon triple point~\cite{Zhou2022-nk}. Electrons were subsequently deposited onto the solid neon surface at base temperature 
	using a filament biased at 0.1 V~\cite{Jennings2024-sb}.

	\begin{acknowledgments}
		We acknowledge Yasunobu Nakamura and the Superconducting Quantum Electronics Research Team and the Semiconductor Science Research Support Team at RIKEN for their support in fabricating the sample. We thank Shuhei Tamate, Yutian Wen, Hayato Goto, Pasquale Scarlino, and Xinhao Li for useful discussions. This work was supported by the JST under the Adopting Sustainable Partnerships for Innovative Research Ecosystem (ASPIRE) program, Grant Number JPMJAP2532.
		This work was also supported by the RIKEN Hakubi Program, the RIKEN Center for Quantum Computing, and the JST FOREST. D.~J. acknowledges support from the Air Force Office of Scientific Research (AFOSR) under Award No.~FA9550-23-1-0636. 
	\end{acknowledgments}
	
	\clearpage
	\section{Supplementary material}
	\setcounter{figure}{0}
	\renewcommand{\thefigure}{S\arabic{figure}}

	\subsection{Finding a suitable electron and dispersive readout of the qubit\label{sec:electron_selection_readout}}
	
	We probe the coupled electron--resonator system by measuring the transmission through the feedline as a function of probe frequency while sweeping the voltages applied to the electrodes.
	
\begin{figure}[ht!]
		\centering
		\includegraphics[width=1\columnwidth]{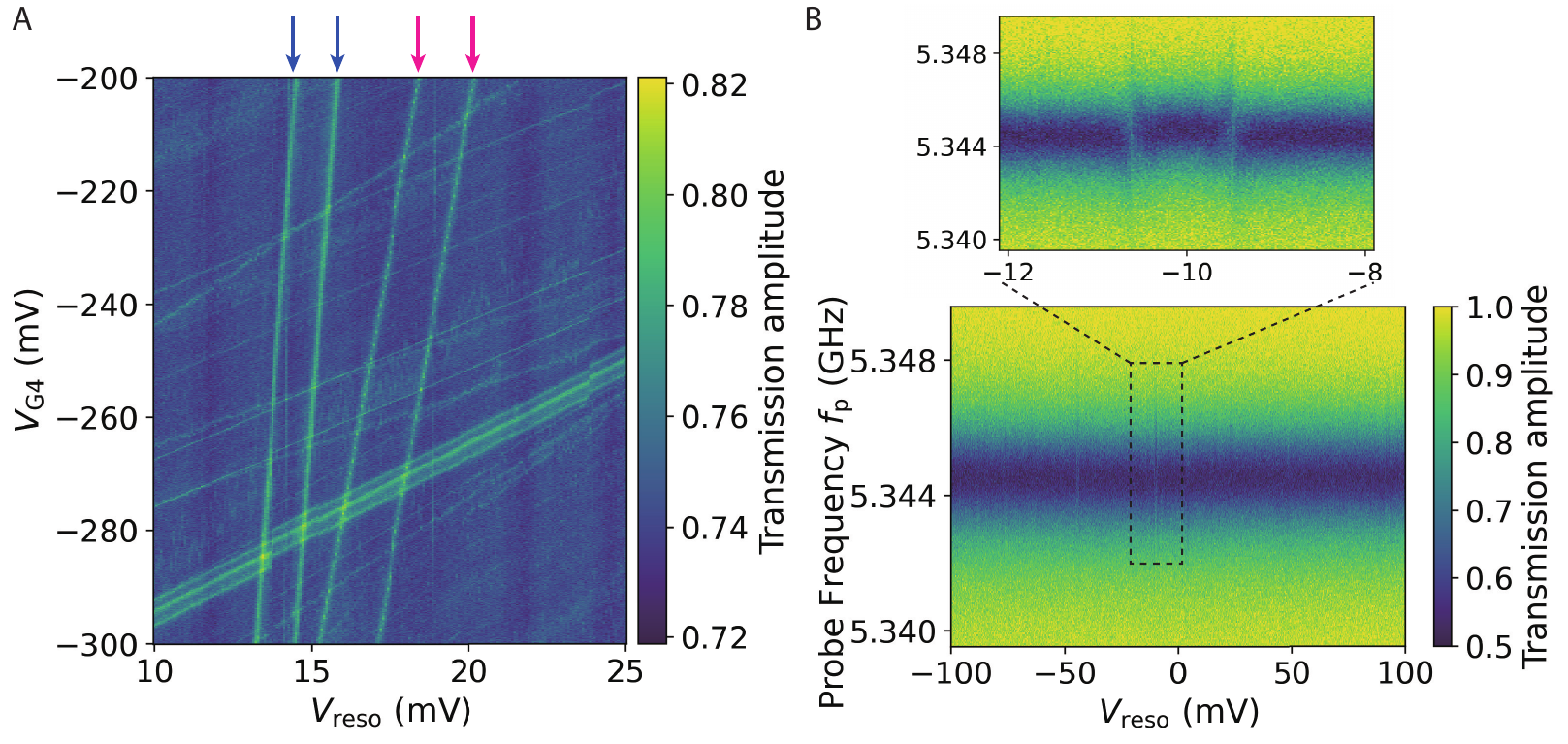}
		\caption{(A) Transmission amplitude as a function of \(V_\mathrm{G4}\) and \(V_\mathrm{reso}\). The probe frequency \(f_p\) is fixed to \(f_r\). Sets of blue and pink arrows indicate pairs of parallel lines that frequently appear in the map. (B) Transmission amplitude as a function of \(f_\mathrm{p}\) and \(V_\mathrm{reso}\) with \(V_\mathrm{G4}=-500\)~mV fixed. Inset: Zoomed-in view focusing on an electron exhibiting particularly strong coupling to the resonator. Note that the measurement conditions in (A) and (B) are different from those in Figs.~\ref{fig1}D--E of the main text. \label{figS1}}
	\end{figure}

	Figure~\ref{figS1}A shows the transmission measured with the probe frequency fixed at \(f_p = f_r\) and applied to the feedline, while sweeping the voltages \(V_{\mathrm{G4}}\) and \(V_{\mathrm{reso}}\). When the electron charge-qubit frequency \(f_q\) becomes resonant with the resonator frequency \(f_r\), the transmission changes, appearing as lines in the voltage map. Two approximately parallel lines are observed because the qubit frequency follows Eqs.~\ref{eq:E} and \ref{eq:alpha_DC} of the main text; therefore, there are two different combinations of \(V_{\mathrm{G4}}\) and \(V_{\mathrm{reso}}\) that satisfy the resonance condition \(f_q = f_r\). Performing similar measurements while sweeping the voltages applied to G1, G2, and G3 did not produce such features in the transmission map. This is because the nanowire resonator acts as a screening barrier: electrons confined inside the resonator region remain strongly coupled to the resonator, while their coupling to the outer electrodes G1, G2, and G3 is significantly suppressed.

	After identifying electrons that are likely to form a suitable qubit from the voltage map, we fix either \(V_{\mathrm{G4}}\) or \(V_{\mathrm{reso}}\), and sweep the other voltage together with the probe frequency \(f_p\). Figure~\ref{figS1}B shows the transmission measured with \(V_{\mathrm{G4}} = -500~\mathrm{mV}\) fixed, while sweeping \(V_{\mathrm{reso}}\) and \(f_p\). An avoided crossing appears when \(f_p = f_r = f_q\), and the size of the avoided crossing reflects the coupling strength between the electron and the resonator. A larger avoided crossing indicates stronger electron--resonator coupling.

	\subsection{Extraction of the resonator–qubit coupling strength \label{sec:extract_g}}
\begin{figure}[ht!]
		\centering
		\includegraphics[width=0.6\columnwidth]{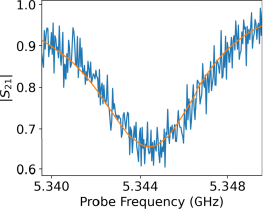}
		\caption{Line cut of Fig.~\ref{fig1}D at $V_{\mathrm{reso}}=-21.18$~mV, after calibrating out phase rotation and cable offsets. The solid line is a fit to Eq.~\ref{eq:S21_ref}. \label{fig:g_facror_fit}}
	\end{figure}

	Figure~\ref{fig:g_facror_fit} shows a line cut of Fig.~\ref{fig1}D taken at 
	$V_{\mathrm{reso}}=-21.18$ mV. We fit this trace using
	\begin{equation}
		|S_\mathrm{21}| = \sqrt{\frac{\tilde{A}^2 + \tilde{B}^2}{A^2 + \tilde{C}^2}},
		\label{eq:S21_ref}
	\end{equation}
	where
	\begin{align}
		\tilde{A} &= (\omega_0-\omega)
		-\frac{g^2(\omega_q-\omega)}
		{(\omega_q-\omega)^2+(\gamma/2)^2},\\
		\tilde{B} &= \frac{\kappa_\mathrm{int}}{2}
		+ \frac{g^2 \gamma/2}
		{(\omega_q-\omega)^2+(\gamma/2)^2},\\
		\tilde{C} &= \frac{\kappa_\mathrm{int}+\kappa_\mathrm{ext}}{2}
		+ \frac{g^2 \gamma/2}
		{(\omega_q-\omega)^2+(\gamma^2/2)}.
	\end{align}
	From this fit, we extract the coupling strength $g = 2.1 \pm 0.2\,\mathrm{MHz}$ 
	and the linewidth $\gamma = 11.7 \pm 2.3\,\mathrm{MHz}$. 
	The extracted linewidth $\gamma \approx 1/T_2^*$ is consistent with the dephasing time measured in the time domain.
	
	\subsection{On-chip DC filter\label{sec:dc_filter_transmission}}
	
\begin{figure}[t]
		\centering
		\includegraphics[width=0.6\columnwidth]{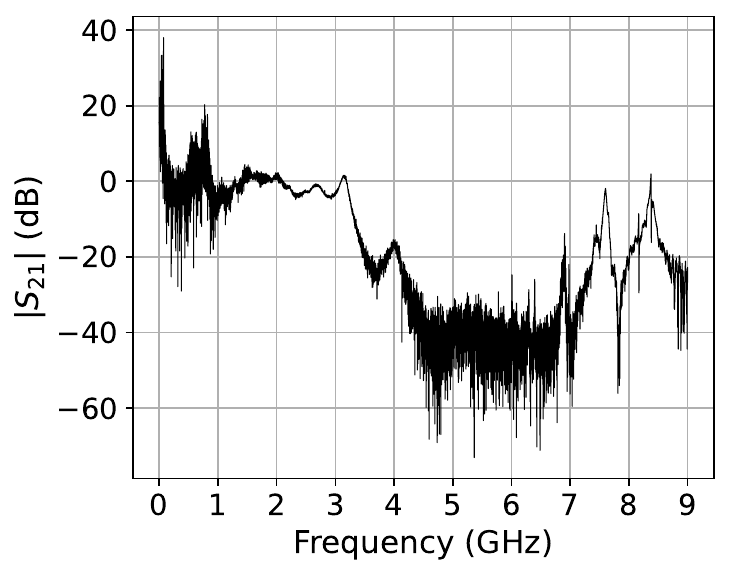}
		\caption{The extrapolated transmission magnitude $|S_{21}|$ of the on-chip DC filter measured at $T=660$ mK for a VNA output power of $-20$ dBm.
			The apparent values above 0 dB are attributed to background subtraction and residual measurement noise.}
		\label{fig:dc_filter_transmission}
	\end{figure}

	To independently evaluate the frequency response of the on-chip DC filter, we fabricated an identical DC filter on a separate chip and measured the transmission magnitude $|S_{21}|$ between 0 and 9 GHz at 660 mK with a VNA output power of -20 dBm (Fig.~\ref{fig:dc_filter_transmission}). Strong attenuation is maintained throughout the 4–8 GHz band, which includes the resonator frequency, demonstrating that the filter operates effectively over the frequency range relevant to our measurements.
	
	\subsection{Electron Position Estimation} \label{sec:position_estimation}
\begin{figure}[!ht]
		\centering
		\includegraphics[width=\columnwidth]{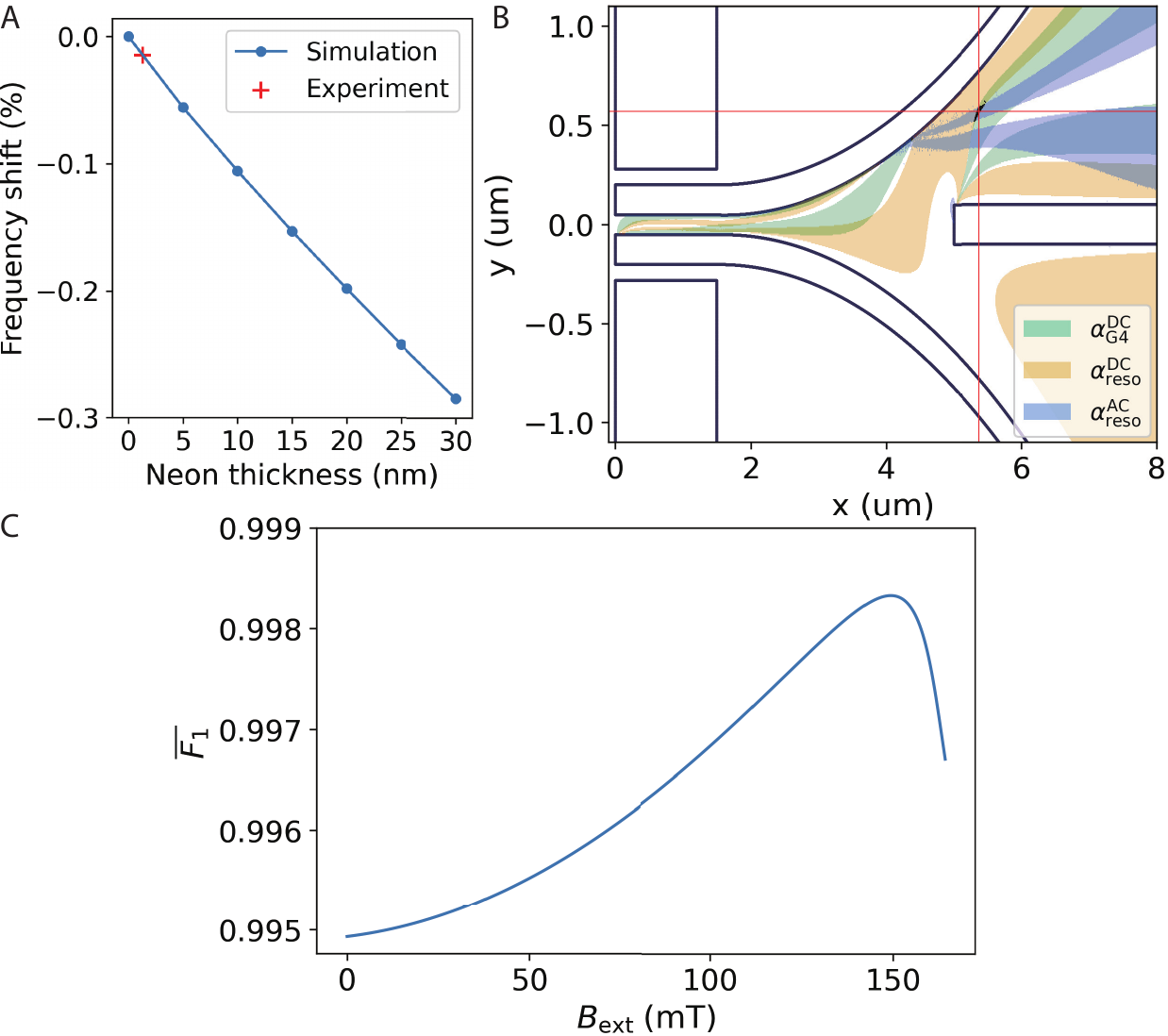}
		\caption{(A) Blue dots: simulated relation between neon thickness and the resonance frequency shift of the superconducting resonator. The red cross corresponds to the experimentally measured resonance frequency shift of $-0.015\%$ after neon deposition. (B) Possible positions of the electron in the slice with $z=2.3$ nm, $d=100$ nm, $\theta=15^\circ$, and $r_\text{tol}=2$. The green, orange, and blue regions indicate the positions where the simulated values of $\alpha^\text{DC}_\text{G4}$, $\alpha^\text{DC}_\text{reso}$, and $\alpha_\mathrm{reso}^\mathrm{AC}$ are within the tolerance range of $r_\text{tol}=2$, respectively. The dark gray region is the overlap area where all simulated parameters are within the tolerance range. (C) Spin qubit gate fidelity versus $B_\text{ext}$, with the electron position fixed at the cross point identified in (B).}\label{fig-S-pos}
	\end{figure}

	The vertical position $z$ of the electron can be estimated from the neon thickness, which is inferred from the resonator frequency shift~\cite{Zhou2022-nk, Tian2025-fz}. We simulate the resonance frequency shift as a function of neon thickness, as shown in Fig.~\ref{fig-S-pos}A. In the experiment, the observed frequency shift of $-0.015\%$ corresponds to a neon thickness of 1.3~nm. Together with the $\sim$1~nm height of the electron above the neon surface~\cite{Kanai2024-bo}, we estimate an electron height of $z=2.3$~nm above the Si substrate, and given the 60~nm etching depth into the silicon, the electron is located 57.7~nm below the bottom of the NbTiN layer. We note that this value corresponds to an average neon thickness inferred from the resonator response. In reality, the local neon thickness at the electron position may deviate from this average value due to spatial inhomogeneity of the neon film, and therefore the estimated electron position should be regarded as an effective value with  uncertainty.
	
	For the horizontal direction, we consider a double-well trap parallel to the surface, i.e., in the $x$-$y$ plane, tilted by an angle $\theta$ from the $x$ axis as shown in Fig.~\ref{fig1}C of the main text. The distance between the expected positions of the local bases $|L\rangle$ and $|R\rangle$ is defined as $d=|\boldsymbol{r}_L-\boldsymbol{r}_R|$, where $\boldsymbol{r}_L=\langle L|\boldsymbol{r}|L\rangle=(x_L,y_L)$ and $\boldsymbol{r}_R=\langle R|\boldsymbol{r}|R\rangle=(x_R,y_R)$ are the positions of the two localized states, respectively. We define the electron position in the $x$-$y$ plane as the midpoint between $\boldsymbol{r}_L$ and $\boldsymbol{r}_R$. The electron couples to the resonator electric field via its electric dipole moment. Since we observe no dependence of the qubit frequency on the voltages applied to G1, G2, and G3, we infer that the electron is located inside the resonator area, where the electric fields from G1, G2, and G3 are screened by the resonator inner conductor.
	
	For the DC field simulation, we apply 1 V to G4 or to the resonator while 
	grounding all other electrodes. The resulting field distribution is then 
	scaled to the specific voltages applied to G4 and the resonator. Since $\hbar\varepsilon=E_L-E_R$ is the difference between onsite energies of local bases $|L\rangle$ and $|R\rangle$, when an DC electric potential $V^\text{DC}(x,y)$ is applied by electrodes, $\varepsilon$ will be changed by $\hbar\delta\varepsilon=eV^\text{DC}(x_L,y_L)-eV^\text{DC}(x_R,y_R)$. Therefore, the DC coupling $\alpha^\text{DC}_{i}$ is evaluated by
	\begin{align}
		\alpha^\text{DC}_{i,\text{sim}}(x,y,z,d,\theta)=\frac{|(V^\text{DC}_i(x_L, y_L)-V^\text{DC}_i(x_R, y_R)|}{1 \text{V}}. 
	\end{align}
	
	For the AC field simulation, since the resonator edges forming the parallel capacitor are much shorter than the total resonator length, we disconnect them from the rest of the resonator and apply $+0.5$~V and $-0.5$~V to the two segments, respectively, while grounding all other electrodes. In this way, we simulate the oscillating electric field at position \((x,y,z)\) for a voltage oscillation amplitude of 1 V across the capacitor plates, and infer the electric field \(\boldsymbol{E}(x,y,z)\). The simulated AC lever arm is then given by
	\begin{align}
		\alpha_\mathrm{reso,\text{sim}}^\mathrm{AC}
		(x,y,z,d,\theta)=
		\frac{\langle e \vert \boldsymbol{r}\cdot \boldsymbol{E}(x,y,z)\vert g\rangle}{1~\mathrm{V}}.
	\end{align}
	
	Since the simulation is based on an ideal case, the simulated value $O_\text{sim}$ will not match the experimental value $O_\text{exp}$ explicitly. We introduce the tolerance  $r_\text{tol}$ to indicate the range of parameter $O$: 
	\begin{align}
		O_\text{exp}/r_\text{tol}<O_\text{sim}<O_\text{exp}r_\text{tol}. 
	\end{align}
	We plot the tolerant areas corresponding to $\alpha^\text{DC}_{\text{G4},\text{sim}}(x,y)$, $\alpha^\text{DC}_{\text{reso},\text{sim}}(x,y)$, and $\alpha^\text{AC}_{\text{reso},\text{sim}}(x,y)$ on a specific slice plane $(z,d,\theta)$, where the overlap region represents the possible electron positions. Such overlap regions exist only for certain combinations of $(z,d,\theta)$. We calculate several slice planes $(d,\theta)$ at $z=2.3$~nm with $d\in\{10,50,100\}$~nm and $\theta\in\{0^\circ,5^\circ,\cdots,90^\circ\}$. The results indicate that $\theta\le15^\circ$ and $d\ge50$~nm for a tolerance of $r_\text{tol}=2$, as summarized in Table~\ref{tab:e-place-F}. The tolerant areas corresponding to $\alpha^\text{DC}_{\text{G4},\text{sim}}(x,y)$, $\alpha^\text{DC}_{\text{reso},\text{sim}}(x,y)$, and $\alpha^\text{AC}_{\text{reso},\text{sim}}(x,y)$ for $d=100$~nm and $\theta=15^\circ$ are shown in Fig.~\ref{fig-S-pos}B. We also performed simulations for $z=7$~nm and determined the possible electron positions and the corresponding spin-qubit fidelities discussed in the next section, finding that the results were qualitatively unchanged. This suggests that variations in the local neon thickness on the order of several nanometers do not significantly affect these results.
	
	\begin{table}[!ht]
		\centering
		\begin{tabular}{r|r|r|r|r|r|r|r}
			\hline
			\shortstack{$\theta$ \\ $(^\circ)$} & \shortstack{$d$ \\ (nm)} & \shortstack{$x$ \\ (nm)} & \shortstack{$y$ \\ (nm)}  & \shortstack{$\overline{F_1}$ \\  ~ } \rule{0pt}{6ex} & \rule{0pt}{5ex}\shortstack{$B_\parallel$ \\ (mT)} & \rule{0pt}{5ex}\shortstack{$\delta B_\perp$ \\ (mT)} & \rule{0pt}{5ex}\shortstack{$B_\text{opt}$ \\ (mT)} \\
			\hline
			0 & 50 & 4.19 & 0.07 & 0.9943 & 101 & 3.5 & 69.6 \\
			0 & 100 & 3.43 & 0.02 & 0.9982 & 149 & 6.7 & 12.6 \\
			5 & 50 & 4.24 & 0.19 & 0.9947 & 104 & 3.7 & 65.9 \\
			5 & 50 & 4.23 & 0.06 & 0.9940 & 98 & 3.4 & 73.0 \\
			5 & 50 & 5.33 & 0.37 & 0.9899 & 21 & 2.8 & 150.9 \\
			5 & 100 & 3.56 & 0.11 & 0.9976 & 145 & 4.9 & 21.8 \\
			5 & 100 & 5.51 & 0.28 & 0.9919 & 15 & 3.5 & 155.9\\
			10 & 50 & 4.38 & 0.33 & 0.9951 & 107 & 3.9 & 62.7 \\
			10 & 50 & 5.21 & 0.46 & 0.9934 & 28 & 4.3 & 140.1 \\
			10 & 100 & 3.89 & 0.25 & 0.9979 & 138 & 6.9 & 23.9 \\
			10 & 100 & 5.46 & 0.43 & 0.9932 & 15 & 4.2 & 154.1 \\
			15 & 50 & 4.86 & 0.54 & 0.9967 & 77 & 7.5 & 82.6 \\
			15 & 100 & 5.37 & 0.57 & 0.9952 & 15 & 6.0 & 148.4 \\
			\hline
		\end{tabular}
		\caption{Possible electron positions $(x,y)$ at slice $(d,\theta)$ within tolerance range $r_\text{tol}=2$. 
			For $d=10$ nm or $\theta\ge20^\circ$, there is no position that simultaneously matches 
			$\alpha^\text{DC}_\text{G4}$, $\alpha^\text{DC}_\text{reso}$, and $\alpha^\text{AC}_\text{reso}$ 
			with the experimental results. 
			$B_\parallel$ is the $y$-component of the magnetic field induced by the cobalt magnet at $(x,y)$. 
			$\delta B_{\perp}$ is the difference between the transverse magnetic fields at the two local 
			minima of the double-well potential. $\overline{F_1}$ is the single-qubit gate fidelity at $B_\mathrm{ext}=0$ as discussed in Sec.~\ref{sec:Prospect_spin} of the main text. The maximum fidelity $\overline{F}_{1,\mathrm{max}}=0.9983$, achieved at $B_\mathrm{ext}=B_\mathrm{opt}$ for each configuration, is unchanged across all listed configurations.}
		\label{tab:e-place-F}
	\end{table}
	
	\subsection{EDSR via Ferromagnet-Induced Spin-Charge Coupling \label{sec:EDSR}}
	In order to realize electric dipole spin resonance (EDSR), we will add cobalt ferromagnets on top of the device used in this work, as described in Fig.~\ref{fig:spin} of the main text. By applying a sufficiently large external magnetic field $B_{\mathrm{ext}}$ along the $y$ axis, the Co layer is magnetized to saturation, with saturation magnetization $M_s = 1.4\times10^6~\mathrm{A/m}$~\cite{braik2021in}. As a result, the $y$ direction becomes the dominant spin quantization axis. The total magnetic field defining the Zeeman splitting in angular frequency is expressed as
	\begin{equation}
		b_{\parallel}= \frac{g_e\mu_B}{\hbar}
		|\boldsymbol{B}\cdot \hat{\mathbf{e}}_y | \approx \frac{g_e\mu_B}{\hbar} (B_\parallel + B_\mathrm{ext}),
	\end{equation}
	where $\hat{\mathbf{e}}_y$ is the unit vector along the $y$ direction, $g_e$ is the free-electron $g$-factor, $\mu_B$ is the Bohr magneton, $\boldsymbol{B}$ is the total magnetic field given by the sum of the stray field $\boldsymbol{B}_{\mathrm{Co}}$ generated by the cobalt layer and the externally applied uniform magnetic field $\boldsymbol{B}_{\mathrm{ext}}$ along the $y$ direction, and $B_\parallel$ is the $y$-component of the stray field induced by the cobalt layer at the electron position.
	The Co layer also gives rise to a stray field difference between the two potential minima, $\delta \boldsymbol{B}=\boldsymbol{B}_R-\boldsymbol{B}_L$, whose transverse component perpendicular to the $y$ direction is
	\begin{equation}
		b_{\perp}= \frac{g_e\mu_B}{\hbar}\sqrt{(\delta \boldsymbol{B}\cdot \hat{\mathbf{e}}_x)^2+(\delta \boldsymbol{B}\cdot \hat{\mathbf{e}}_z)^2} = \frac{g_e\mu_B}{\hbar} \delta B_{\perp},
	\end{equation}
	where $\hat{\mathbf{e}}_x$ and $\hat{\mathbf{e}}_z$ are the unit vectors along the $x$ and $z$ directions, respectively. This transverse field difference $b_{\perp}$ hybridizes the charge and spin degrees of freedom and thereby enables EDSR. The single-qubit gate fidelity $\overline{F_1}$ (Eq.~\ref{eq:single_qubit_fidelity}) depends on $\Lambda$, which represents the spin-charge coupling strength, and $\Lambda$ can be tuned via the external magnetic field $B_\text{ext}$. Figure~\ref{fig-S-pos}C shows the dependence of $\overline{F_1}$ on $B_\text{ext}$ for the electron position discussed in the main text ($(x,y)=(5.37,0.57)$, $z=2.3$~nm, $d=100$~nm, $\theta=15^\circ$). $\overline{F_1}$ reaches its maximum of $\overline{F}_{1,\mathrm{max}}=0.9983$ at $B_\text{ext} =148.4$~mT. We estimate the spin single-qubit gate fidelity for the candidate electron positions listed in Table~\ref{tab:e-place-F}.

	\subsection{Two possible mechanisms for charge qubit frequency shift versus driving power}
	
	Here we use the notation $\omega_r = 2\pi f_r$, $\omega_d = 2\pi f_d$, $\omega_q = 2\pi f_q$, 
	and $\Omega = 2\pi f_\text{Rabi}$.
	
	\subsubsection{Dispersive coupling \label{sec:dispersive}}
	
	In the dispersive regime, i.e., $g \ll \Delta_{qr}$ with $\Delta_{qr}=\omega_q-\omega_r$ being the qubit--resonator detuning, the interaction can be treated perturbatively.  In the rotating frame of the driving frequency $\omega_d $, the Hamiltonian 
	is given by~\cite{Blais2004-ty}:
	\begin{align}
		H=~&(\omega_r-\omega_d) a^\dagger a+\frac{1}{2}\left(\omega_q-\omega_d+\frac{g^2(2a^\dagger a+1)}{\Delta_{qr}}\right)\sigma_z\nonumber\\
		~&+\varepsilon_d(a^\dagger +a) +\frac{g\varepsilon_d }{\Delta_{qr}}\sigma_x.
	\end{align}
	Here, $\varepsilon_d=\frac{A}{A_0}\varepsilon_{d,0}$ is the resonator drive amplitude. The last term defines the Rabi frequency, $\Omega=2g\varepsilon_d/\Delta_{qr}$ and from this we can obtain $\varepsilon_{d,0}=2\pi\times5.86$~GHz corresponds to the full drive amplitude (10~dBm source power). The term $2g^2a^\dagger a/\Delta_{qr}$ is a qubit frequency shift 
	proportional to the photon number operator. Due to the driving, the average photon number is proportional to the square of the driving amplitude, $\varepsilon_d$, according to
	\begin{align}
		\langle a^\dagger a\rangle=\frac{\varepsilon_d^2}{(\omega_r-\omega_d)^2+\kappa^2/4}. 
	\end{align}
	Neglecting the cavity dissipation rate $\kappa$ with respect to the detuning, $\kappa\ll |\omega_r-\omega_d|$
	the qubit frequency shift can be rewritten as
	\begin{align}
		\frac{2g^2\langle a^\dagger a\rangle}{\Delta_{qr}}= \frac{2g^2\varepsilon_d^2}{(\omega_r-\omega_d)^2\Delta_{qr}}=
		\frac{\Omega^2 \Delta_{qr}}{2(\omega_r-\omega_d)^2}
		\approx\frac{\Omega^2}{2\Delta_{qr}}. 
	\end{align}
	With $\Omega/2\pi = f_\text{Rabi}=76$ MHz and $\Delta_{qr}/2\pi=-324$ MHz as measured experimentally, 
	the estimated frequency shift is 9.1 MHz, whereas the experimentally observed 
	frequency shift is 21 MHz.

	\subsubsection{Ponderomotive effect \label{sec:Ponderomotive}}
	
	The ponderomotive effect is generated by an inhomogeneous microwave driving field acting on a moving charged particle. Since we assume that the electron moves only along the $x'$ axis connecting the two potential minima, the relevant spatial dependence reduces to the one-dimensional coordinate $x'$. Here, $E(x')$ denotes the magnitude of the electric field evaluated along the $x'$ axis~\cite{Kibble1966-ph,cook1985pra,Bucksbaum1987-pd}:
	\begin{align}
		U_\text{p}(x')
		=
		\frac{e^2}{4m\omega_d^2}E^2(x'). \label{eq:U_p}
	\end{align}
	
\begin{figure}[!ht]
		\centering
		\includegraphics[width=\columnwidth]{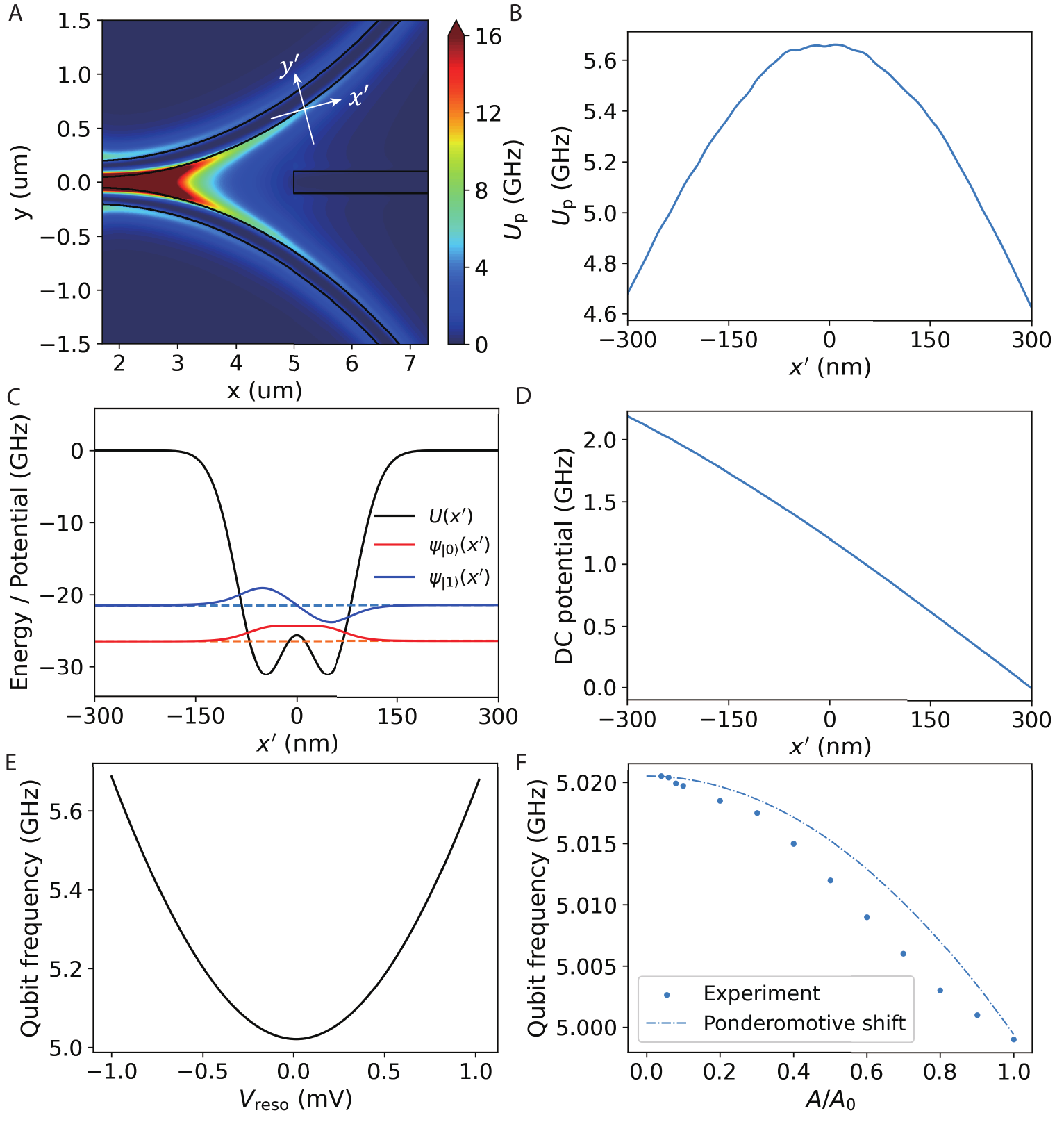}
		\caption{Ponderomotive-drive-induced qubit-frequency shift. (A) The ponderomotive potential in the $(x,y)$ plane at 40~nm above the top surface of the NbTiN electrodes. The white cross indicates the center position of the double-well trap considered here. The coordinate $x'$ denotes the axis connecting the two wells, with an angle of $\theta=15^\circ$ between $x'$ and $x$.
			(B) The ponderomotive potential $U_p(x')$ along the $x'$ direction for $A/A_0=1$, which is maximized at the center of the double well.
			(C) The double-well potential along $x'$ at $\varepsilon=0$ without driving. $\psi_{|0\rangle}(x')$ and $\psi_{|1\rangle}(x')$ are the wavefunctions of the ground and first excited orbital states. The distance between the two local minima is 100~nm.
			(D) Simulated DC potential induced by applying a $+0.1$~mV DC voltage to the resonator while all other electrodes are grounded.
			(E) Qubit frequency tuned by the DC voltage applied to the resonator without driving. In this work, qubit experiments are performed at $\varepsilon=0$, corresponding to the minimum of this parabola. Note that in the experiment, $\varepsilon=0$ corresponds to $V_\mathrm{reso}=-21.21$~mV as shown in Fig.~\ref{fig1}E, while in the simulation this offset is incorporated into the static double--well potential.
			(F) Qubit frequency as a function of normalized drive amplitude $A/A_0$. The dash-dotted line is the simulated qubit frequency at $\varepsilon=0$ for each drive amplitude. The blue dots are experimental results.}\label{fig-S-ponder}
	\end{figure}

	To quantitatively reproduce the experimentally observed charge qubit frequency shift 
	(Fig.~\ref{fig4}C) via the ponderomotive effect, the electron position must be located at the representative position marked by the white cross in Fig.~\ref{fig-S-ponder}A, which lies at the edge of the resonator with an assumed $z$ position 40 nm above the top surface of the NbTiN layer ($z = 120$ nm). This vertical position differs from the electron position deduced in Sec.~\ref{sec:position_estimation}; however, it cannot be excluded given that the neon surface roughness could locally elevate the electron above the averaged height, with the horizontal electron  position being closer to the resonator edge. The ponderomotive potential along the $x'$ axis, $U_\mathrm{p}(x')$ (Eq.~\ref{eq:U_p}), for this location is shown in Fig.~\ref{fig-S-ponder}B.
	
	We obtained the electron wavefunctions by solving the Schrödinger equation for the static double-well potential at $\varepsilon = 0$, as shown in Fig.~\ref{fig-S-ponder}C, using parameters chosen to reproduce the experimentally observed qubit frequency. When a voltage is applied to the resonator, an additional potential is introduced. Figure~\ref{fig-S-ponder}D shows the case when $+0.1$~mV is applied to the resonator. As shown in Fig.~\ref{fig-S-ponder}E, the qubit frequency follows a parabolic dependence as a function of $V_\mathrm{reso}$. The qubit frequency at $\varepsilon=0$ matches the qubit frequency measured under weak driving conditions. When the ponderomotive potential (Fig.~\ref{fig-S-ponder}B) is added to the static double-well potential at $\varepsilon=0$ (Fig.~\ref{fig-S-ponder}C), the barrier between the two wells increases, resulting in a decrease in the qubit frequency $\Delta/2\pi$. The strength of the ponderomotive potential scales with the drive amplitude. Consequently, the induced change in the qubit frequency also scales with the drive amplitude, as seen in Fig.~\ref{fig-S-ponder}F. Note that Fig.~\ref{fig-S-ponder}F represents the ponderomotive shift alone, while the the dash-dotted line in Fig.~\ref{fig4}C reflects the combined contribution of the ponderomotive and dispersive shifts.

%

\end{document}